\documentclass{aa}  

\usepackage{graphicx,url,rotating}
\usepackage{txfonts,color}
\newcommand{\teff}{$T_{\rm eff}$}
\newcommand{\vmic}{$V_{\mathrm{mic}}$}
\newcommand{\msun}{M$_\odot$}

\begin{document} 

   \title{Rare Find: Discovery and chemo-dynamical properties of two s-process enhanced RR Lyrae stars}

   \author{V. D'Orazi
          \inst{1,2,3,9}
          \and
           G. Iorio\inst{4}
            \and
          B. Cseh\inst{5,6}
          \and
    C. Sneden\inst{3}
        \and
       H. Abdollahi \inst{5,7}
           \and
            L. Moln\'ar\inst{5,8}
            \and
         A. Bobrick\inst{9,10}
            \and
     G. Bono\inst{1,11}
           \and
      V. F. Braga\inst{11}
            \and
           A. Karakas\inst{9}
           \and
            M. Lugaro\inst{5,8,9}
            \and
            S. W. Campbell\inst{9}
         \and
           M. Fabrizio\inst{11,12}
           \and
           G. Fiorentino\inst{11}
            \and
           I. U. Roederer\inst{13}
           \and
           N. Storm\inst{14}
           \and
           M. Tantalo\inst{11}
           \and
           J. Crestani\inst{1,11}
           }

   \institute{Department of Physics, University of Rome Tor Vergata, via della Ricerca Scientifica 1, 00133, Rome, Italy\\ \email{vdorazi@roma2.infn.it}
         \and
        INAF Osservatorio Astronomico di Padova, vicolo dell'Osservatorio 5, 35122, Padova, Italy
        \and
        Department of Astronomy \& McDonald Observatory, The University of Texas at Austin, 2515 Speedway, Austin, TX 78712, USA
        \and
        Departament de F\'isica Qu\`antica i Astrof\'isica, Universitat de Barcelona, c. Mart\'i i Franqu\`es, 1, 08028 Barcelona, Spain 
        \and
        Konkoly Observatory, HUN-REN Research Centre for Astronomy and Earth Sciences, MTA Centre for Excellence, Konkoly Thege Mikl\'os \'ut 15-17, Budapest 1121, Hungary
        \and
        MTA-ELTE Lend{\"u}let "Momentum" Milky Way Research Group, Szent Imre h. u. 112., 9700, Szombathely, Hungary
        \and
        School of Astronomy, Institute for Research in Fundamental Sciences (IPM), Tehran, 19568-36613, Iran
        \and
        E\"otv\"os Lor\'and University, Institute of Physics and Astronomy, H-1117 P\'azm\'any P\'eter s\'et\'any 1/A, Budapest, Hungary
        \and
         School of Physics and Astronomy, Monash University, Clayton, VIC 3800, Australia
        \and
        OzGrav: Australian Research Council Centre of Excellence for Gravitational Wave Discovery, Clayton, VIC 3800, Australia
        \and
        INAF -- Osservatorio Astronomico di Roma, via Frascati 33, Monte Porzio Catone, Italy
        \and
        Space Science Data Center – ASI, Via del Politecnico SNC, 00133 Rome, Italy
        \and
        Department of Physics, North Carolina State University, Raleigh, NC 27695, USA
        \and
        Max-Planck-Institut fur Astronomie, Konigstuhl 17, D-69117 Heidelberg, Germany
      }

   \date{Received 2025; accepted ...}

% \abstract{}{}{}{}{} 
% 5 {} token are mandatory
 
  \abstract
  {}
   {We report the serendipitous discovery of two RR Lyrae stars exhibiting significant s-process element enrichment, a rare class previously represented solely by TY Gruis. Our goal is to characterise these objects chemically and dynamically, exploring their origins and evolutionary histories.}
   {Using high-resolution spectroscopy from HERMES@AAT and UVES@VLT, we derived detailed chemical abundances of key s-process elements (Y, Ba, La, Ce, Nd, Eu), carbon along with $\alpha$-elements (Ca, Mg, Ti). We also employed Gaia DR3 astrometric data to analyse their kinematics, orbital properties, and classify their Galactic population membership. We compared observational results with theoretical asymptotic giant branch nucleosynthesis models to interpret their enrichment patterns.}
   {Both stars exhibit clear signatures of s-process enrichment, with significant overabundances in second-peak elements such as Ba and La compared to first-peak Y and Zr. Comparison with AGB nucleosynthesis models suggests their progenitors experienced pollution of s-process-rich material, consistent with early binary interactions. However, notable discrepancies in dilution factors highlight the need for more refined low-metallicity asymptotic giant branch (AGB) models.
    We also explore and discuss alternative scenarios, including sub-luminous post-AGB-like evolution or double episodes of mass transfer. In the latter case, the star initially undergoes a mass transfer when it is on the main sequence, accreting material from a former AGB companion. Subsequently, as the star evolves along the red giant branch, it may again transfer mass to its companion, before becoming an RR Lyrae star. }
   {Our findings confirm the existence of s-process-enhanced RR Lyrae stars and demonstrate the importance of combining chemical and dynamical diagnostics to unveil their complex evolutionary pathways. Future detailed binary evolution modelling and long-term orbital monitoring are essential to resolve their formation scenarios and assess the role of binarity in the evolution of pulsating variables.}

   \keywords{Stars: variables: RR Lyrae -- Stars: abundances -- Stars: AGB and post-AGB -- Stars: binaries: general.
}

   \maketitle
%
%-------------------------------------------------------------------

\section{Introduction}\label{sec:intro}

 Stellar nucleosynthesis involves multiple pathways operating across a range of stellar masses, synthesising different elements. Light metals (from carbon to zinc) are primarily produced via charged-particle fusion within stellar cores and shells. In contrast, the formation of elements beyond the iron peak (Fe-group) is dominated by neutron-capture (n-capture) processes, which are classified based on the neutron capture timescale relative to radioactive decay: the slow (s-) process, which predominantly occurs in asymptotic giant branch (AGB) stars, and the rapid (r-) process, associated with highly energetic astrophysical sites such as type {\sc ii} supernovae and neutron star mergers\citep{burbidge1957,sneden2008,cowan2021,lugaro2023}.

Evolved stars with unusual abundance patterns in very light/heavy elements serve as natural laboratories for studying nucleosynthesis pathways. Among these, Barium (Ba) stars \citep{bidelman1951} are a prominent class characterised by strong spectral lines of s-process \footnote{Neutron-capture elements up to bismuth (Z = 83) can be synthesised in various amounts by both the r- and s-processes, while heavier elements are purely r-process products.  Also, the production of some elements is so dominated by either the r-process (e.g., Eu, Gd, Dy, Pt) or the s-process (e.g., Sr, Y, Zr, Ba, La, Ce) that they are commonly referred to as r- or s-process elements.  We will follow that convention in this paper. Moreover, alongside the s- and r- neutron capture processes, an intermediate neutron capture process (i-process) is thought to exist \citep{hampel2019,choplin2022}, but it will not be addressed here.} elements like Sr, Y, Ba, La, and others. They are generally G- or K-type dwarfs and giants with effective temperatures between approximately \teff = 4,000 -- 6,000 K. Their peculiar chemical abundance patterns (for their evolutionary state), coupled with radial velocity variations indicating binarity, support the hypothesis that Ba stars have accreted s-process enriched material from former AGB companions, now white dwarfs, in binary systems \citep{mcclure1980,jorissen1998}. This mass transfer scenario explains the surface enrichment despite the stars’ evolutionary stages being unrelated to the production of such elements internally.

Detailed spectroscopic analyses, using high-resolution and high signal-to-noise observations, have revealed significant correlations between s-process element abundances and metallicity, as well as insights into the nucleosynthetic conditions in their progenitor AGB stars. Studies comparing observed abundance ratios (such as [hs/ls], between heavier and lighter s-process elements\footnote{The s-process elements, formed via neutron capture in stars, are mainly concentrated in two peaks on the periodic table: the first at neutron magic number N=50, which includes elements such as Sr, Y, and Zr, and the second at N=82, comprising elements like Ba, La, Ce, Pr, and Nd (see e.g., \citealt{lugaro2012} and references therein). Elements in the first peak are commonly referred to as light s-process elements (ls), while those in the second peak, such as Ba, La, and Ce, are classified as heavy s-process elements (hs).}) with theoretical models support low-mass ($\approx$2–4 M$_\odot$) AGB stars as the primary sites of s-process nucleosynthesis involved in Ba star formation \citep{decastro2016,cseh2018,cseh2022,roriz2021,roriz2024,yang2024}. Furthermore, the orbital properties of Ba systems, including their period-eccentricity distributions, have been extensively studied to better constrain models of binary evolution and mass transfer \citep{jorissen2016,jorissen2019,krynski2025}. Beyond classical Ba stars, related objects such as CH stars and carbon-enhanced metal-poor (CEMP) stars extend this paradigm to different stellar populations. CH stars, similar to Ba stars but generally of lower metallicity, are also members of binary systems where s-process enhancement results from past mass transfer during the AGB phase of a companion \citep{keenan1942, mcclure1990}. CEMP stars, particularly the CEMP-s subclass, show large carbon enhancements ([C/Fe] $> +0.7$), accompanied by s-process element overabundances, and are typically very metal-poor stars. The binary nature of many CEMP-s stars has been well established, supporting their formation via mass transfer processes akin to those in Ba and CH stars \citep{lucatello2005, abate2015, hansen2016,jorissen2016}. Understanding the properties of these chemically peculiar stars not only sheds light on the nucleosynthetic processes within AGB stars but also probes the formation and evolution of binary and multiple star systems across different stellar populations. 

While Ba/CH/CEMP-s stars are observed among giants and unevolved stars (occurrence rate $\lesssim 1\%)$, TY Gruis \citep{preston2006} stood as the sole identified s-process-enhanced, carbon-rich RR Lyrae (RRL). Low-mass stars undergoing helium burning on the HR diagram horizontal branch will cross the instability strip (IS), becoming RR Lyrae pulsating variables during this transition \citep[and references therein]{catelan2004}. Their defining characteristics include short-period brightness variations (hours to less than a day) and pulsations in the fundamental (RRab), first overtone (RRc), or both modes simultaneously (RRd) \citep{catelan2007, sos2011}. The astrophysical significance of RRLs stems from their prevalence in old stellar populations \citep{iben1974,preston2019}, well-established period-luminosity relations enabling precise distance determination \citep{longmore1986, bono2011,braga2021}, and readily distinguishable light curve morphologies \citep{bailey1902,jurcsik1996, smolec2005}. 
RRL stars display a broad spectrum of chemical compositions, from metallicities as low as 1/1000 of solar to solar. Moreover, while metal-poor and extremely metal-poor RRL stars are characterised by $\alpha$-enhancements \citep{hansen2011, dorazi2025}, metal-rich RRLs exhibit subsolar $\alpha$-element abundances \citep{chadid2017, dorazi2024}. This finding has motivated the development of alternative evolutionary pathways, such as binary interactions \citep[e.g.,][]{karczmarek2017,bobrick2024}. 

\citet{preston2006} conducted an extensive examination of the metal-poor ([Fe/H] $\approx -2.0 \pm 0.2$) RRL TY Gruis, discovering substantial carbon and neutron-capture element excesses suggestive of mass transfer from a former AGB companion. Unfortunately, the presence of the Blazhko effect complicated radial velocity measurements, hindering the search for orbital evidence of a binary system.

This paper presents the serendipitous discovery of two additional s-process-rich RR Lyrae stars, expanding the known population beyond the single previous detection by \citet{preston2006} and extending it towards higher metallicities. The paper is structured as follows: Section \ref{sec:sample} details the sample selection, observational methods, and abundance analysis procedures. In Section \ref{sec:results} we present the key findings, and Section \ref{sec:discussion} provides a comprehensive discussion of the results.

%--------------------------------------------------------------------
\section{Sample, observations and abundance analysis}\label{sec:sample}

\begin{figure*}
   \centering
   \includegraphics[width=0.9\textwidth]{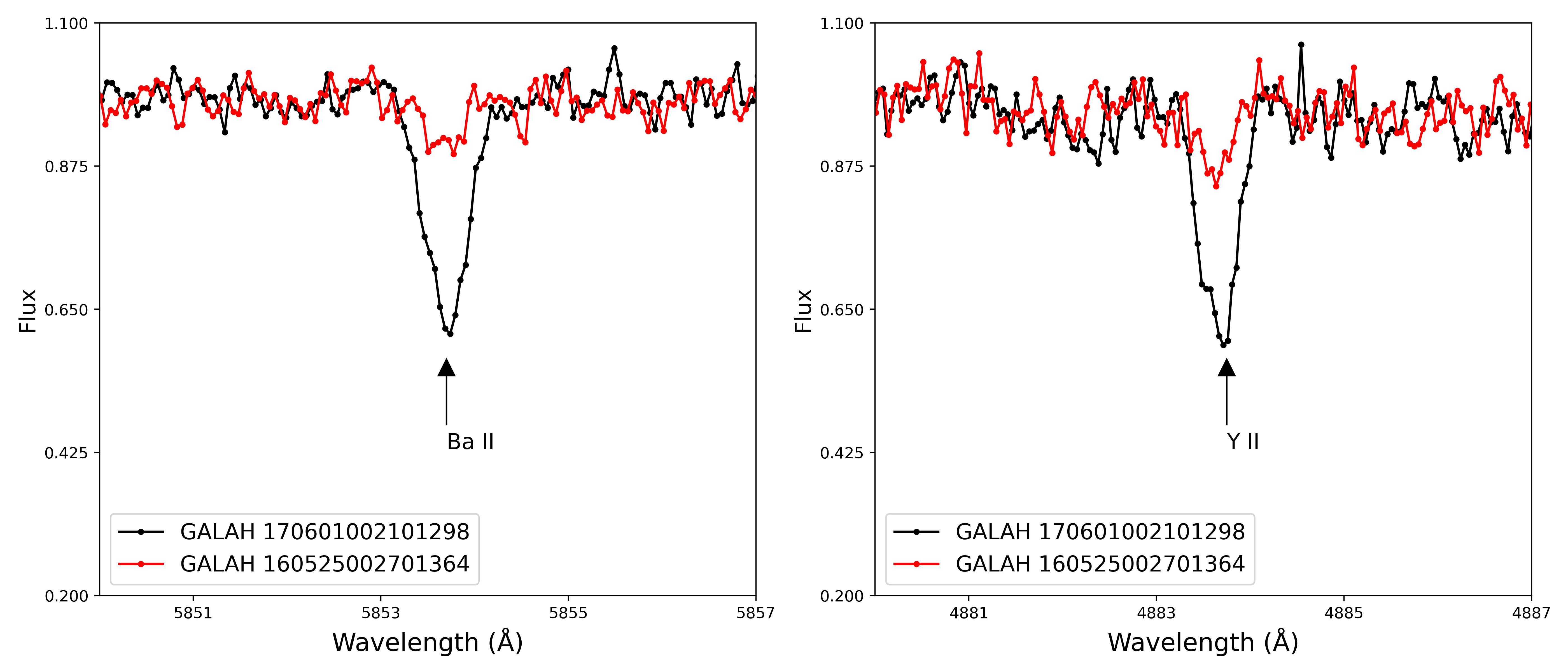}
      \caption{Comparison of the observed spectra from GALAH DR4 for our sample RRL DQ Hya (GALAH 170601002101298) and another RRL with very similar atmospheric parameters and metallicity by \cite{dorazi2024}.}\label{fig:spectral_comparison}
   \end{figure*}

We cross-referenced our newly updated proprietary variable star catalogue, featuring over 300,000 RR Lyrae stars (Braga et al., in prep.), with the GALAH DR4 results. This large-scale survey is conducted using the HERMES spectrograph \citep{sheinis2015} at the Australian Astronomical Telescope (AAT), which operates at a nominal resolution of R = 28,000 and includes one million stars \citep{desilva2015, buder2025}. This cross-referencing resulted in 470 objects identified as RR Lyrae stars. We downloaded the GALAH stacked spectra corresponding to these objects, each with a maximum exposure time of 3600 seconds, and examined them for indications of s-process enrichment through the synthesis of the Ba II and Y II lines (see Section \ref{sec:intro}). Initially, we adopted stellar parameters from GALAH DR4 \citep{buder2025}. We classified abundances as s-process-rich if they exceeded [s/Fe] > 0.3 dex (see also \citealt{cseh2018}). However, our new abundance determinations for Ba and Y led to the exclusion of many Ba-rich candidates previously reported in the official DR4 catalogue, as they were not confirmed by our analysis. This discrepancy is likely due to suboptimal assumptions in the automatic abundance determinations for variable stars; for instance, GALAH DR4 claims an overabundance of Ba in many of these stars, but not of Y, La. From our original sample, we retained 12 candidates with confirmed s-process enhancement, most of which exhibit metallicity above [Fe/H] $\gtrsim -0.5$. Given the possibility of misclassification between $\delta$ Scuti and RRL variables at these relatively high metallicities, our subsequent step was to search for simultaneous carbon enhancement. To achieve this, we carried out spectral synthesis calculations of the C I line at 6587 \AA, considering stars with [C/Fe] < 0 as not enhanced and excluding them from our sample. As demonstrated by \cite{sneden2018}, significant overabundances of s-process elements are common in $\delta$ Scuti stars, yet carbon is not enhanced, as expected from the mass transfer scenario of former AGB companions. Consequently, after confirming significant enhancements in Ba and Y (i.e., [Ba,Y/Fe] $\gtrsim 0.3$), we examined whether these stars also showed concurrent, substantial carbon enrichment. At these evolutionary stages, carbon is generally expected to be depleted due to the first dredge-up and possible extra-mixing processes. Therefore, a [C/Fe] $\gtrsim 0$ can be regarded as an indication of anomalously high carbon abundances. This refined selection resulted in identifying only two s-process-rich RRL stars: DQ Hya and ASAS J153830-6906.4. Comparison of the observed HERMES spectra for DQ Hya and a standard RRL from our previous paper \citep{dorazi2024} is given in Figure \ref{fig:spectral_comparison}.

Spectroscopic observations of our s-rich RRLs were performed using HERMES@AAT as part of the GALAH survey. Due to the short pulsation period of RRLs, typically less than a day, we chose not to use the survey spectrum because it consists of stacked spectra, leading to total exposure times exceeding one hour. Instead, we retrieved single-exposure spectra from the AAT archive and combined only those with total exposure times less than 40 minutes. This approach enables us to achieve a minimum signal-to-noise ratio (SNR) of 30 per pixel in the red arm (approximately 6500 Å) for both spectra. Additionally, one of the two stars, DQ Hya, was previously observed with UVES at the VLT, featuring a resolving power of R $\approx$ 35,000. We selected two spectra, in red and blue, acquired at the pulsation phase around $\phi \approx 0.5$ (see Table \ref{tab:log}), and discarded spectra observed at phases near $\phi$ $\sim$ 0.8–0.9. At these extreme phases, RR Lyrae stars become hotter than 7000 K, making abundance analysis challenging due to the lack of suitable spectral lines. According to \cite{for2011}, the optimal phase range for maximising line identification in the spectra of variable RR Lyrae stars is approximately $\phi$ = 0.3–0.4. While the red spectrum ($\lambda\lambda$ 4800–6800 \AA) has a signal-to-noise ratio (SNR) per pixel of approximately 50, the blue spectrum ($\lambda\lambda$ 3300–4500 \AA) has a lower SNR of 23. To our knowledge, no other high-resolution spectra are available for the other sample star (ASAS J153830-6906.4). 

Table \ref{tab:log} provides details of our spectroscopic observations. We kept the original archival nomenclature for the names of the spectra to facilitate easy retrieval from the corresponding sources. Alternative names for the stars are listed in Column 2 of the table, along with the SNR, nominal resolutions, Modified Julian Date, phase of the spectra and instantaneous radial velocities. We employed \texttt{iSpec} \citep{blanco2019} to compute the radial velocities (RVs) for each spectrum, using a cross-correlation technique with a synthetic spectrum calculated based on the atmospheric parameters listed in the original GALAH DR4, while accounting for the different instrumental resolutions. Our final RV measurements agree very well with the GALAH values, with a mean difference of less than 0.5 km/s. The same set of synthetic spectra was also used to perform continuum normalisation.
As previously done in \cite{dorazi2024,dorazi2025}, we used the MARCS grids of model atmospheres \citep{gustafsson2008} and \texttt{TSFitPy} \citep{gerber2023,storm2023}, the \texttt{Python} wrapper of \texttt{Turbospectrum v.20} \citep[and references therein]{plez2025}. This allows the computation of Non-Local Thermodynamic Equilibrium (NLTE) synthetic profiles on-the-fly and fast optimisation algorithms.

Effective temperatures (\teff)  were determined by fitting the wings of the H$\alpha$ profiles after masking the central region. We employed three different fitting procedures: fitting the entire profile (from 6550 to 6570 \AA) and fitting the left and right wings separately. The measurements from these methods showed good agreement, and we adopted the standard deviation of the three as a conservative estimate of the error in \teff. Surface gravity ($\log g$) was derived using the ionisation balance of neutral and singly 
ionised lines of Fe and Ti. This approach was chosen due to the limited number of available lines in the GALAH spectra, caused by incomplete spectral coverage across the optical band. Similarly, the microturbulent velocity (\vmic) was obtained by coupling lines of Fe and Ti and adopting the same procedure described in \cite{dorazi2024}. 
As this is an iterative process, an initial estimate of the titanium abundance was adopted based on the first-guess measurements of Ca and Mg abundances. The analysis was then repeated iteratively until convergence was achieved between the derived stellar parameters and the Ti abundances. Uncertainties in the surface gravity (log g) were estimated by iteratively adjusting its value until the ionisation equilibrium condition was no longer satisfied within 2.5$\sigma$ of the measurement uncertainty. For the \vmic values, we adopted a conservative error estimate of 0.5 km s$^{-1}$.

The line list used is provided in Table \ref{tab:linelist} and includes references from the Gaia-ESO line list \citep{heiter2021}, the VALD database \citep{vald}, and \cite{lawler2013} for Ti I lines.
From the GALAH spectra, we were able to determine the abundances of certain elements using specific spectral lines: carbon through the C I line at 6587.6 \AA, yttrium using Y II at 4883 \AA, and barium via the Ba II lines at 5853 and 6496 \AA. In contrast, the high-resolution UVES spectra for DQ Hya also allowed us to identify zirconium, lanthanum, cerium, neodymium, and europium, in addition to yttrium and barium (based on a larger set of spectral features). This was possible due to a larger wavelength coverage (see Table \ref{tab:linelist} for the complete linelist). To further corroborate the kinematic classification of our sample stars as belonging to the thin disc, thick disc, or halo of the Galaxy, we also aimed to estimate their $\alpha$-element abundances. We derived calcium and magnesium abundances for both stars.

Examples of spectral synthesis computation for the HERMES and UVES spectra of DQ Hya are presented in Figures \ref{fig:synthesis_carbon} and \ref{fig:synthesis}. NLTE abundances were inferred for Fe, Ca, Mg, Ti, Y, Ba, and Eu, while for the remaining species (C, La, Ce, Nd, Zr), we had to rely on LTE assumptions since departure grids are not available in \texttt{TSFitPy}. As supported by results from \cite{alexeeva2015}, the C I line at 6587 \AA~is not significantly affected by NLTE departures; this is also clarified in the caption of Figure \ref{fig:synthesis}. Moreover, the agreement among the s-process elements—aligning with theoretical expectations due to their common production mechanism—indicates that NLTE effects are likely negligible for La, Ce, and Nd. NLTE corrections for Eu, Ba, and Y are within 0.10 dex, roughly matching our typical uncertainties. Due to the large overabundances observed, these corrections do not influence the main conclusions of this study.

There are two types of uncertainties in abundance determinations: one affecting individual spectral lines, which includes random errors in best-fit determination (or equivalent width measurements), oscillator strengths, damping constants, and continuum displacements, and another affecting the entire set of lines, primarily originating from uncertainties in the stellar atmospheric parameters. For the first type, we use the standard deviation of the mean abundance for each species ($\sigma$), which is reported as the associated error in Tables \ref{tab:hermes_abundances} and \ref{tab:uves_abundances}. Additionally, we evaluated abundance sensitivities by varying one stellar parameter at a time and assessing the resulting impact on the [X/Fe] ratios; sensitivities for star DQ Hya are given in Table \ref{tab:sensitivities}.

\begin{table*}[htbp]
    \centering
        \caption{Log of observations used in this work.}
    \begin{tabular}{lcccccr}
    \hline
    \hline
    Spectrum & Star & SNR & R & MJD & Phase & RV \\
             &      &     &    &    &       & (km s$^{^-1}$)\\
    \hline
       &                    &   HERMES at AAT & & & & \\
       GALAH 170601002101298 & DQ Hya    & 37 & 28 000 & 57905.37247933 & 0.67 & $43.0\pm0.3$ \\
       GALAH 140312004501064 & ASAS J153830-6906.4 & 42 & 28 000 & 56728.75554266 & 0.48 & $-18.7\pm0.5$ \\
       &                    &  UVES at VLT & & & & \\
    ADP.2021-10-04T12:15:10.955 & DQ Hya & 52 & 34 540 & 52368.13695321 & 0.54  & $46.2\pm0.3$ \\
    ADP.2021-10-04T12:15:10.969 & DQ Hya & 23 & 36 840 & 52368.13694807 & 0.54 & $45.0\pm0.3$\\
    \hline
    \hline
    \end{tabular}
    \label{tab:log}
\end{table*}

%----------------------------------------------------------------- 
%---------

\section{Results}\label{sec:results}
Stellar parameters and abundances derived from HERMES spectra for DQ Hya and ASAS J153830-6906.4 are summarised in Table \ref{tab:hermes_abundances}. Despite their advanced evolutionary stage (for which C is expected to be depleted by the first dredge-up), both stars display significant enhancements in [C/Fe], approximately $\approx$ +0.5 dex. Their $\alpha$-element abundance patterns are consistent with those of older Galactic populations, such as the thick disc or halo (see Section \ref{sec:chemodyn}). Despite having similar metallicities around [Fe/H] $\approx -$1, DQ Hya exhibits markedly higher enhancements in the [Y/Fe] and [Ba/Fe] ratios: more than an order of magnitude above solar ([Y/Fe]=1.20$\pm$0.10, [Ba/Fe]=1.50$\pm$0.12). In contrast, ASAS J153830-6906.4 shows more moderate values of [Y/Fe] = +0.33$\pm$0.10 and [Ba/Fe] = +0.78$\pm$0.06. In both cases, the second-peak s-process element Ba exceeds the first-peak elements, resulting in positive [Ba/Y] ratios. Due to the limited number of heavy element lines accessible from the GALAH spectra, a more detailed analysis of the s-process pattern was performed using the UVES spectra for DQ Hya (see Table \ref{tab:uves_abundances}). Along with Ba and Y, we derived abundances for the first-peak element (Zr), the second-peak elements (La, Ce, Nd) and the r-process element Eu. Our results confirm a general overabundance of all s-process elements, with second-peak elements (Ba, La, Ce) being more enhanced than the first-peak elements Y and Zr. 

\begin{figure}
   \centering
   \includegraphics[width=0.45\textwidth]{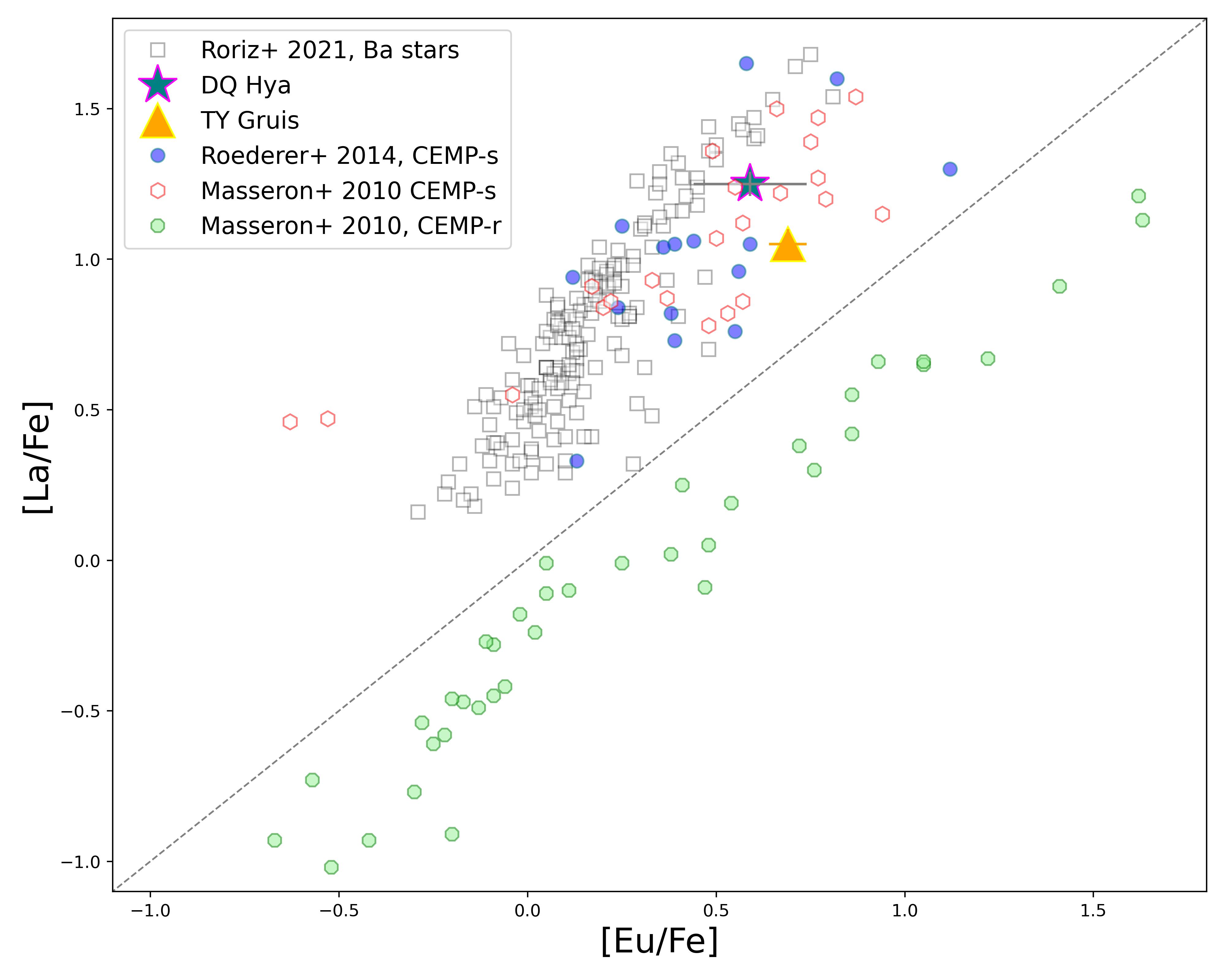}
      \caption{[Eu/Fe] vs. [La/Fe] ratios for CEMP-r and CEMP-s stars by \cite{masseron2010}, CEMP-s stars by \cite{roederer2014}, Ba stars by \cite{roriz2021}, and the s-rich RRLs DQ Hya and TY Gruis. The dashed line is the 1:1 relationship.}\label{fig:eufe_lafe}
   \end{figure}

The [Eu/Fe] ratio also exceeds solar values; however, the observed trend of [Eu/Fe] versus [La/Fe] does not conform to the patterns seen in CEMP-r stars (see e.g., \citealt{masseron2010, karink2021}).
In Figure \ref{fig:eufe_lafe}, we display the relation between [La/Fe] and [Eu/Fe] for various metal-poor heavy-element rich stellar populations: CEMP-s stars from \cite{masseron2010} and \cite{roederer2014}, CEMP-r stars from \cite{masseron2010}, and our enriched RRLs, DQ Hya and TY Gruis, for which Eu abundances could be measured.
These results support classifying these stars as s-process-enhanced objects, likely resulting from mass transfer from a former AGB companion.

\begin{figure}
   \centering
   \includegraphics[width=0.45\textwidth]{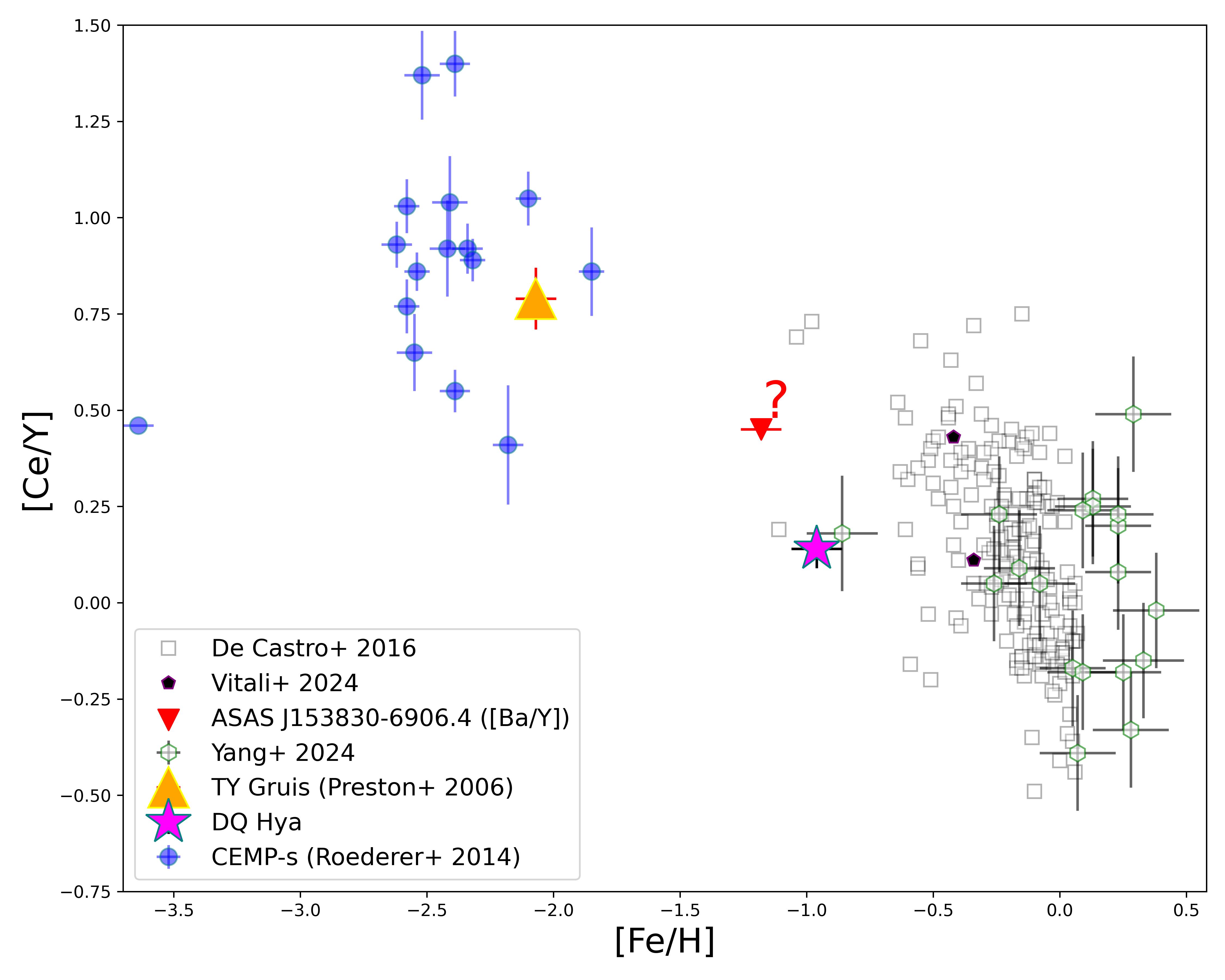}
      \caption{[Ce/Y] ratios as a function of metallicity [Fe/H].}\label{fig:cey_feh}
   \end{figure}

As extensively discussed in \cite{cseh2018,cseh2022}, comparing [hs/ls]\footnote{The ratio between second-peak heavy (hs) and first-peak light (ls) s-process elements is routinely indicated as [hs/ls].} ratios across different studies is challenging, primarily because different studies often include or exclude certain elements. Crucially, Ba abundances are usually considered unreliable due to the strong and nearly saturated behaviour of the commonly used Ba II lines (e.g., 5853, 6496 \AA). Therefore, barium should not be included in the calculation of mean abundances for second-peak s-process elements. Our findings support this approach; in fact, the [Ba/Fe] ratios are systematically higher than those of other second-peak elements such as La and Ce, despite the same nucleosynthetic origin. For this reason, we focus instead on the ratio of two specific elements: yttrium for the first peak and cerium for the second peak, following \cite{cseh2018,cseh2022}. It is important to note that, in the subsequent discussion, Ba is used as a proxy for Ce for the star ASAS J153830-6906.4. Since we were unable to measure Ce abundances from the HERMES spectrum, we adopt [Ba/Y] as equivalent to [Ce/Y] in our analysis. However, as previously mentioned, because of the saturation of Ba II lines, this ratio is likely somewhat larger than [Ce/Y], with an estimated difference of approximately 0.2 dex.
In Figure \ref{fig:cey_feh}, we compare the [Ce/Y] ratios as a function of metallicity for the three s-rich RRLs along with recent literature estimates, namely CEMP-s stars from \cite{roederer2014} and Ba stars from \cite{decastro2016}, \cite{yang2024} and \cite{vitali2024}.
In the metal-poor regime, RRL TY Gruis is located close to the average distribution of CEMP-s stars as reported by \cite{roederer2014}, although it is important to note that these stars exhibit significant scatter. The star shows a moderate enhancement in [Y/Fe] = 0.26 $\pm$ 0.05 and substantial enrichment in second-peak s-process elements, with [Ce/Fe] = 1.05$\pm$0.15 and a [Ce/Y] ratio of 0.79$\pm$0.10. Within this metallicity range ([Fe/H] $\leq -1.5$), all the stars tend to exhibit large [Ce/Y] ratios ($\gtrsim +0.5$), confirming the trend toward producing heavier s-process elements at lower metallicities. This behaviour aligns with the well-known characteristic of the s-process reactions, where a higher neutron-to-seed ratio favours the formation of heavier elements over lighter ones \citep[and references therein]{gallino1998, kobayashi2020}.
DQ Hya and ASAS J153830-6906.4 occupy a pivotal metallicity range, situated between the CEMP-s stars and their higher-metallicity counterparts, i.e., the Ba stars. As observed in both \cite{decastro2016} and \cite{yang2024}, most of their stars are located at metallicities above [Fe/H] $\approx -0.5$. Although based on low number statistics (only six s-rich stars), at around [Fe/H] $\approx-1$, the [Ce/Y] ratios vary from slightly over-solar values to significantly enhanced levels, ranging from [Ce/Y] $\sim 0.15 - 0.75$. Notably, at nearly any metallicity bin, there is a considerable spread in [Ce/Y] ratios, suggesting that this variation may be intrinsic rather than due to observational uncertainties.\footnote{We note that there is an offset between the [Ce/Y] ratios reported by \cite{decastro2016} and \cite{yang2024}, with the former tending to show higher [Ce/Y] at similar metallicities. The origin of this discrepancy remains unclear, especially as the Yang et al. sample appears to be at slightly higher metallicities. However, exploring this difference is beyond the scope of this paper, as it is not directly relevant to our focus on the lower metallicity RRLs.}
Finally, we include in Figure \ref{fig:cey_feh} the two Ba stars recently fully characterised by \cite{vitali2024}, which reside within the metal-rich regime of the metallicity distribution. This study is particularly significant because, in addition to radial velocity variations confirming their binarity, the authors detected a near-UV excess in the spectral energy distribution (SED) of one of their two sample stars, J04034842+1551272 (see their Figure 7). 
To identify similar excesses as potential evidence of a white dwarf (WD) companion, we searched for GALEX NUV \citep{wright2010} photometry for our sample stars. This information is available only for DQ Hya, for which the SED was constructed using the \texttt{VOSA} tool\footnote{\url{https://svo2.cab.inta-csic.es/theory/vosa/}}\citep{bayo2008}. The comparison with a Kurucz model \citep{castelli2003}, as shown in Figure \ref{fig:sed_dqhya}, indicates excellent agreement, with no excess detected in the NUV domain. However, the absence of any near-UV excess does not constitute definitive evidence, particularly given the old age of our RRLs (see Section \ref{sec:formation}), where the WD companion may simply be too faint to be detected.

\begin{table}
    \centering
    \setlength{\tabcolsep}{3pt} 
        \caption{Stellar parameters and abundances inferred from the HERMES/GALAH spectra. The reported uncertainties represent only the random errors.}\label{tab:hermes_abundances}
        \begin{tabular}{lcc}
            \hline
Star      &  DQ Hya  &  {\small ASAS J153830-6906.4} \\
\teff (K) &  6200$\pm$80  &  6093$\pm$70  \\
$\log g$  &   2.80$\pm$0.25  &  2.45$\pm$0.25  \\ \relax
 [Fe/H]$_{\rm NLTE}$  &  $-0.95\pm0.10$  &  $-1.18\pm0.10$ \\
\vmic (km/s) &  3.0$\pm$0.3  &  3.3$\pm$0.3  \\
$V_{\rm mac}$(km/s)  &  17.0$\pm$2.0  &  12.5$\pm$1.5  \\ \relax
[C/Fe]$_{\rm LTE}$    &  0.50$\pm$0.20  &  0.50$\pm$0.20  \\ \relax
[Ca/Fe]$_{\rm NLTE}$  &  0.18$\pm$0.10  &  0.25$\pm$0.08  \\ \relax
[Mg/Fe]$_{\rm NLTE}$  &  0.25$\pm$0.10  &  0.20$\pm$0.10  \\ \relax
[Ti/Fe]$_{\rm NLTE}$  &  0.40$\pm$0.15  &  0.42$\pm$0.06  \\ \relax
[Y/Fe]$_{\rm NLTE}$   &  1.20$\pm$0.10  &  0.33$\pm$0.10  \\ \relax
[Ba/Fe]$_{\rm NLTE}$  &  1.50 $\pm$0.12 &  0.78$\pm$0.06  \\ 
            \hline
        \end{tabular}
\end{table}

\begin{table}[ht]
    \centering
    \caption{Stellar parameters and additional n-capture element abundances of DQ Hya from UVES spectra.}
    \label{tab:uves_abundances}
    \begin{tabular}{lr}
        \hline
        Star & DQ Hya (UVES) \\
        \teff (K) & 6150$\pm$80\\[6pt]
        $\log g$ & 2.50$\pm$0.20\\[6pt]
        [Fe/H]$_{\rm NLTE}$ & $-0.97 \pm 0.11$\\[6pt]
        \vmic (km/s) & 3.00$\pm$0.25\\[6pt]
        $V_{\rm mac}$ (km/s) & 13.6$\pm$1.5\\[6pt]
        [Y/Fe]$_{\rm NLTE}$ & 1.02$\pm$0.07\\[6pt]
        [Zr/Fe]$_{\rm LTE}$ & 0.95$\pm$0.08\\[6pt]
        [Ba/Fe]$_{\rm NLTE}$ & 1.60$\pm$0.06\\[6pt]
        [La/Fe]$_{\rm LTE}$ & 1.25$\pm$0.04\\[6pt]
        [Ce/Fe]$_{\rm LTE}$ & 1.16$\pm$0.10\\[6pt]
        [Nd/Fe]$_{\rm LTE}$ & 0.89$\pm$0.11\\[6pt]
        [Eu/Fe]$_{\rm NLTE}$ & 0.59$\pm$0.15\\ 
        \hline
    \end{tabular}
\end{table}

%----------------------------------------------------------------- 
%-----------------------------------------------------------------
%----------------------------------------------------------------- 
%-----------------------------------------------------------------

\section{Discussion}\label{sec:discussion}

\subsection{Pulsation properties and absolute magnitude} \label{sec:puls}

DQ Hya and ASAS J153830–6906.4 display typical RRab light curves, with some distinct characteristics.
DQ Hya, with a period of  $\approx0.4$ days and an amplitude of  $\approx1.1$ G-mag, belongs to the group of HASP (high amplitude short period) RR Lyrae \citep{fiorentino2015,belokurov2018}. 
ASAS J153830–6906.4 shows a longer period ($\approx0.6$ days) and lower amplitude ($\approx0.6$ G-mag).
Despite these differences, the light-curve properties imply similar photometric metallicities, $[{\rm Fe/H}]_\mathrm{photo} \approx -1.0 \pm 0.3$\footnote{We derived this value using the empirical relation in \cite{iorio2021}; similar values are obtained with the relations in \cite{dekany2022} and \cite{li2023}.}, consistent with the spectroscopic measurements in Table~\ref{tab:hermes_abundances}.
These properties are typical of field RR Lyrae at intermediate metallicity.

From Gaia parallaxes\footnote{Given the high parallax-over-error ratio for both stars ($> 10$), distances are estimated by simple inversion.} and reddening-corrected magnitudes (following \citealt{iorio2021}), we obtain $M_\mathrm{G} \approx 0.7$ for ASAS J153830–6906.4 ($\log L/\mathrm{L_\odot}\approx 1.6$) and $M_\mathrm{G} \approx 0.0$ for DQ Hya ($\log L/\mathrm{L_\odot}\approx 1.9$).
Applying the parallax offset ($\approx -0.03$~mas) from \cite{garofalo2022} yields $M_\mathrm{G} \approx 0.9$ ($\log L/\mathrm{L_\odot}\approx 1.5$) and $M_\mathrm{G} \approx 0.2$ ($\log L/\mathrm{L_\odot}\approx 1.8$), respectively.

Such a large difference is unexpected given the existence of a theoretical and empirical $M_G$–[Fe/H] relations in the optical bands \citep[see, e.g.,][]{marconi2022,garofalo2022}.
In particular, from the \cite{garofalo2022} relation calibrated on Gaia DR3 RR Lyrae, we would expect $M_\mathrm{G} \approx 0.7$ for both stars.
For ASAS J153830–6906.4, the difference with respect to this prediction is modest and within the combined formal and systematic uncertainties ($\approx 0.2$~mag), but for DQ Hya the discrepancy is significant.
Two scenarios are possible:
(1) DQ Hya is a genuine outlier in the $M_G$–[Fe/H] relation, potentially indicating evolutionary effects (HB stars increases their luminosity evolving toward the AGB phase), peculiar properties or a distinct formation channel \citep[see, e.g.,][ and Section~\ref{sec:formation}]{marconi2018} or
(2) the Gaia parallax is biased, possibly due to blending or unresolved binarity (Section~\ref{sec:astro}).
In the latter case, adopting the empirical $M_G$–[Fe/H] relation of \cite{garofalo2022}, we obtain a distance from the Sun $d \approx 2.9\ \mathrm{kpc}$ (in agreement with the distance obtained from the MIR P-L relationship), compared to $d \approx 4.3\ \mathrm{kpc}$ from parallax inversion (or $d \approx 3.8\ \mathrm{kpc}$ with the offset). In Figure \ref{fig:HD} we show the Kiel diagram for our 2 sample s-rich RRLs along with the PARSEC evolutionary tracks \citep{chen2015} and the instability strip location from \cite{marconi2015}.

\subsection{Kinematics and chemodynamics}\label{sec:chemodyn}

\begin{figure}
    \centering
    \includegraphics[width=1.0\linewidth, trim=30 0 50 0, clip]{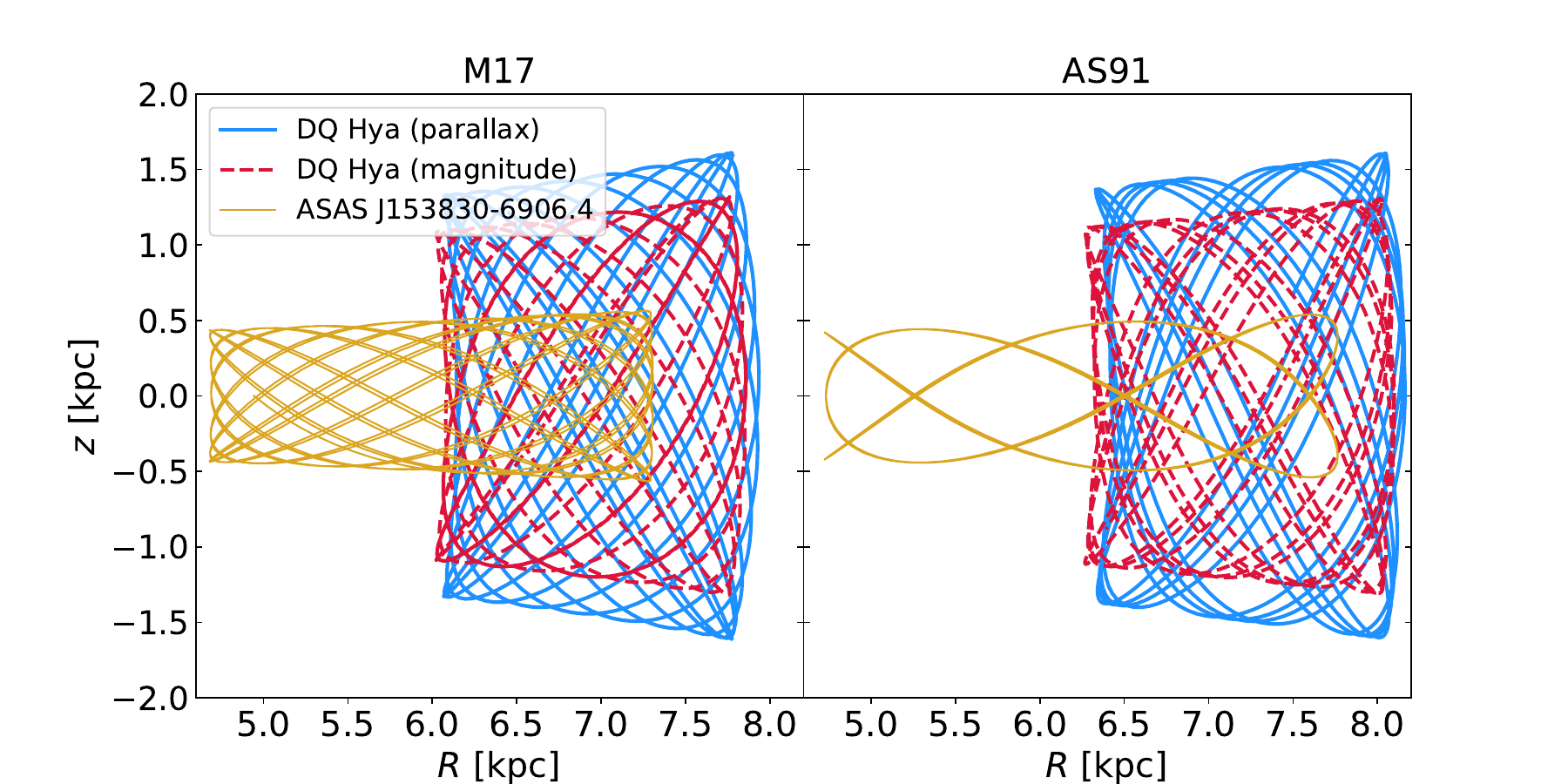}
    \caption{Integrated orbits of DQ Hya (thick solid line: distance from parallax inversion; dashed line: distance estimated from the RR Lyrae magnitude–metallicity relation) and ASAS J153830–6906.4 (thin solid line: distance from parallax inversion), computed using the Milky Way potential models by \cite{mcmillan2017} (M17, left panel) and \cite{allen1991} (AS91, right panel).}
    \label{fig:orbit}
\end{figure}

\begin{figure}
    \centering
    \includegraphics[width=0.9\linewidth]{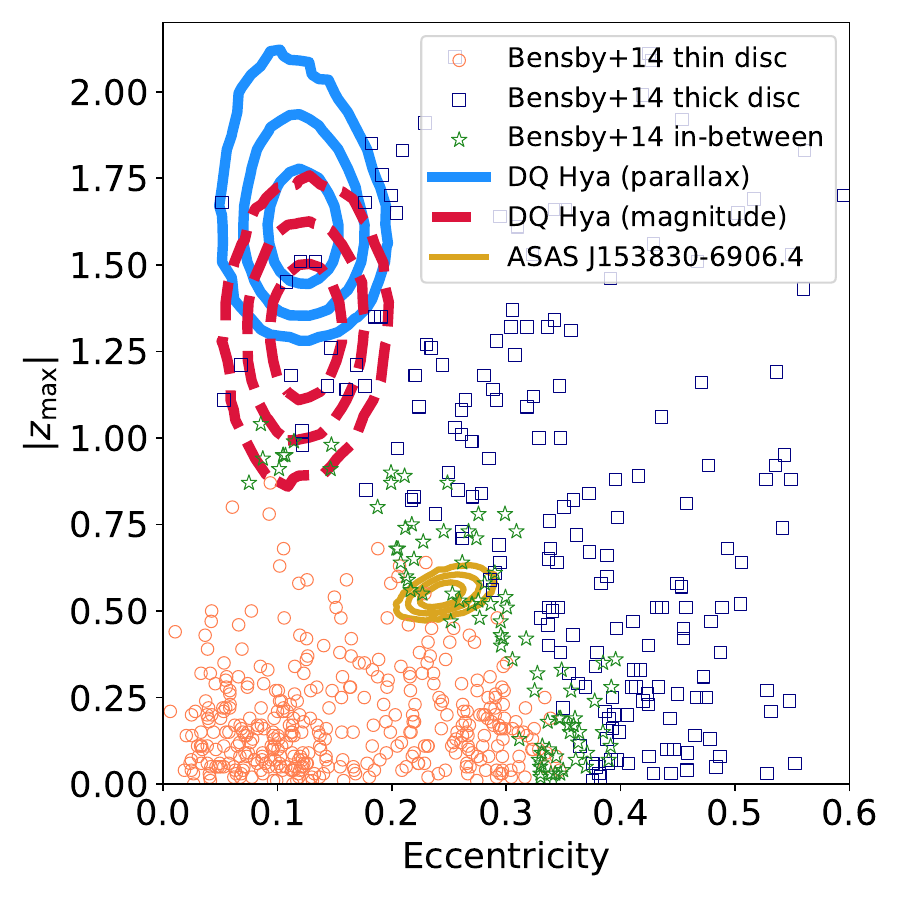}
    \caption{Comparison of the orbital properties, eccentricity and maximum distance from the Galactic plane, of DQ Hya and ASAS J153830–6906.4 with the sample of disc stars from \cite{bensby2014} (B14).
    The contours represent the 68\%, 95\%, and 99.7\% confidence intervals from $10^5$ orbit realisations; line styles and colours follow those used in Figure~\ref{fig:orbit}.
    Circle and square markers indicate stars classified by B14 as likely thin disc and thick disc members, respectively, based on a thick-to-thin disc likelihood ratio of $<0.5$ and $>2$. Star-shaped markers represent stars in between the two components with intermediate likelihood ratios.
    Both our orbit integrations and those in B14 adopt the Galactic potential from \cite{allen1991}.}
    \label{fig:bcomp}
\end{figure}

To infer the kinematic properties of the two stars, we use their Gaia DR3 astrometric data (parallax, proper motions, and radial velocity) to initialise orbit integrations in the Milky Way, adopting two models for the Galactic potential: that of \citet{mcmillan2017} (M17) and \citet{allen1991} (AS91).
For the orbit integration, we use the leapfrog integrator implemented in the orbit module of the \textsc{Python} package \textsc{Galpy} \citep{bovy2015}\footnote{\url{http://github.com/jobovy/galpy}, we used version 1.11.0. For the M17 potential, we use the built-in model {\sc McMillan17}, while for AS91 we implement a custom potential module based on the built-in {\sc Irrgang13I} model, which shares the same dynamical components as AS91.}.  All the orbits are integrated forward for 1.5 Gyr. To transform the observed quantities into six-dimensional phase-space coordinates in the Galactic reference frame, we use the \textsc{Python} module \textsc{Astropy}\footnote{\url{https://github.com/astropy/astropy}, we used the version 7.1.0} \citep{astropy2013, astropy2018, astropy2022}.  
For the M17 potential, we adopt a solar position of $R_\odot = 8.21$ kpc and a circular velocity at the solar radius of $V_\mathrm{lsr} = 233.1\ \mathrm{km\ s^{-1}}$; for AS91, we use $R_\odot = 8.5$ kpc and $V_\mathrm{lsr} = 220\ \mathrm{km\ s^{-1}}$. In both cases, we assume that the Sun is located 20.8 pc above the Galactic midplane \citep{bennett2019}. The Sun’s peculiar motion with respect to the local standard of rest (LSR) is taken as  
$(U_\odot, V_\odot, W_\odot) = (11.1 \pm 1.3,\ 12.2 \pm 2.1,\ 7.25 \pm 0.6)\ \mathrm{km\ s^{-1}}$ \citep{schonrich2010},  
where the right-handed Cartesian Galactic coordinate system is defined such that $U$ points toward the Galactic center, $V$ in the direction of Galactic rotation, and $W$ toward the North Galactic Pole.

We estimate the distances of the two stars from the Sun by inverting their parallaxes, an approach justified by their small relative parallax uncertainties (less than 10\%).  
In the case of DQ Hya, however, the discrepancy between the parallax-based distance and the one expected from its absolute magnitude given its metallicity suggests a possible bias in the parallax measurement (see Section~\ref{sec:puls}).  
Therefore, for this star, we repeat the orbit integration using the distance inferred from the magnitude--metallicity relation for field RR~Lyrae stars \citep{garofalo2022}.
To account for uncertainties, in addition to the orbit computed using the fiducial parameters, we integrate $10^5$ additional orbits by sampling the posterior distribution of the astrometric parameters.
We assume normally distributed errors and take into account both the uncertainties and covariances reported in the Gaia DR3 catalogue.
Additionally, we sample the values of the Sun’s peculiar motion by considering the errors reported in \cite{schonrich2010}.
We also repeat the orbit integrations by applying the parallax offset estimated for field RR Lyrae stars by \citet{garofalo2022} (approximately $-0.03$~mas), as well as the offset determined as a function of magnitude, colour, and position by \citet{lindegren2021} (approximately $-0.04$~mas for DQ Hya and $-0.03$ ASAS J153830–6906.4).
The results of the orbit integrations are consistent with those obtained without applying any parallax correction; therefore, we do not discuss them further in the rest of the paper.

Figure \ref{fig:orbit} shows the fiducial orbits of the two stars in the Galactic $R$-$z$ plane and Figure \ref{fig:bcomp} shows the joint posterior distribution of the orbit eccentricity and the maximum vertical displacement reached by the orbits; additional properties are summarised in  Table \ref{tab:orbitmemb}.
As illustrated in Figure \ref{fig:orbit}, the orbits computed under the two Galactic potentials are very similar. Therefore, we will discuss them in general terms without focusing on the details of each specific model.
The two stars exhibit tube-like orbits typical of stars belonging to the Galactic disc. 
DQ Hya has a low eccentricity ($\approx 0.1$), but a significant vertical excursion ($z_\mathrm{max}>1$ kpc), instead ASAS J153830–6906.4 remains close to the disc ($z_\mathrm{max}<0.5$ kpc) but exhibits higher eccentricity  ($\approx 0.25$) with notable radial excursion bringing it in the inner part of the Galaxy ($R\lessapprox 5$ kpc). 
These orbital properties put the stars in an intermediate regime between the thick disc stars and the hottest (and likely oldest) thin disc stars. 
Indeed, Figure \ref{fig:bcomp} shows that when compared with the disc stellar catalogue by \cite{bensby2014}, ASAS J153830–6906.4 occupies the regions of stars classified as in-between the likely thin and thick disc stars. DQ Hya seems to be more consistent with the thick disc component, especially considering the orbit in which the distance is estimated from the parallax. To obtain a more quantitative estimate of the thin/thick disc membership likelihood, we adopt a method similar to that of \citet{bensby2014}, but using the thin and thick disc Galactic components as defined by \citet{robin2023}. 
The membership likelihood for each Galactic component is
\begin{equation}
p_i =\frac{\rho_i(\vec{x}) \times
\ \mathcal{N}(\vec{v}|\vec{\theta}_\mathrm{kin,i} )
\ \times  \mathcal{N}([Fe/H]| \vec{\theta}_\mathrm{[Fe/H],i})}{\sum^{N_\mathrm{components}}_i p_i},
\label{eq:member}
\end{equation}
where \( \rho \) is the local volumetric density of the Galactic component at the position of the star (Table~3 in \citealt{robin2023}). The second term is a multivariate normal distribution, 
where $\theta_\mathrm{kin}$ refers to the kinematic properties of the component (velocity dispersion and asymmetric drift, as given in Table~4 of \citealt{robin2023}). The third term is a normal distribution in metallicity, with $\theta_\mathrm{[Fe/H]}$  denoting the mean and standard deviation of the metallicity distribution of the given component as reported in Table~2 of \citealt{robin2023}.
The main difference with respect to the membership analysis performed by \citet{bensby2014} is that we include metallicity in the likelihood computation.  
The likelihood is evaluated for all \( 10^5 \) orbit realisations, and we also sample \( 10^5 \) values of the stellar metallicity, assuming Gaussian uncertainties based on the values reported in Table~\ref{tab:hermes_abundances}.

The results of the membership analysis are reported in Table~\ref{tab:orbitmemb}.  
If the metallicity term in Equation~\ref{eq:member} is neglected, the results are consistent with the qualitative comparison shown in Figure~\ref{fig:bcomp}, suggesting an ambiguous classification between the thick disc and the older components of the thin disc.  
However, once metallicity is included, the membership probability strongly favours an association with the thick disc ($p_\mathrm{thick} > 0.9$) for both stars, regardless of the assumed distance for DQ~Hya or the Galactic dynamical model used. Notably, both stars exhibit enhancement in $\alpha$-elements (Ca, Mg, and Ti), confirming their likely affiliation with the thick disc. This demonstrates that combining kinematic and detailed chemical information is essential; studies relying solely on dynamical parameters without considering elemental abundances can lead to misleading conclusions. We plan to further explore this point in a forthcoming publication. 

%\giuc{@Valentina: I think this is a good point to also include the $\alpha$-element abundances, which are even more constraining than metallicity alone. This would also strengthen the conclusion of the section, highlighting that the combination of kinematics and chemical abundances supports the classification of these stars as genuine thick disc members. After this analysis I am starting to think that the use of the circularity parameter could be a bit misleading, at least in the contest of the thin/thick disc classification  and without a control sample to check against. For this reason I have not discussed it and I think that this could be also removed from Table 4, so that the Table could only store the membership likelihood (since eccentricity and zmax are already shown in Fig. 6), }

\subsection{Comparison with AGB models}

\begin{figure}
   \centering
   \includegraphics[width=0.5\textwidth]{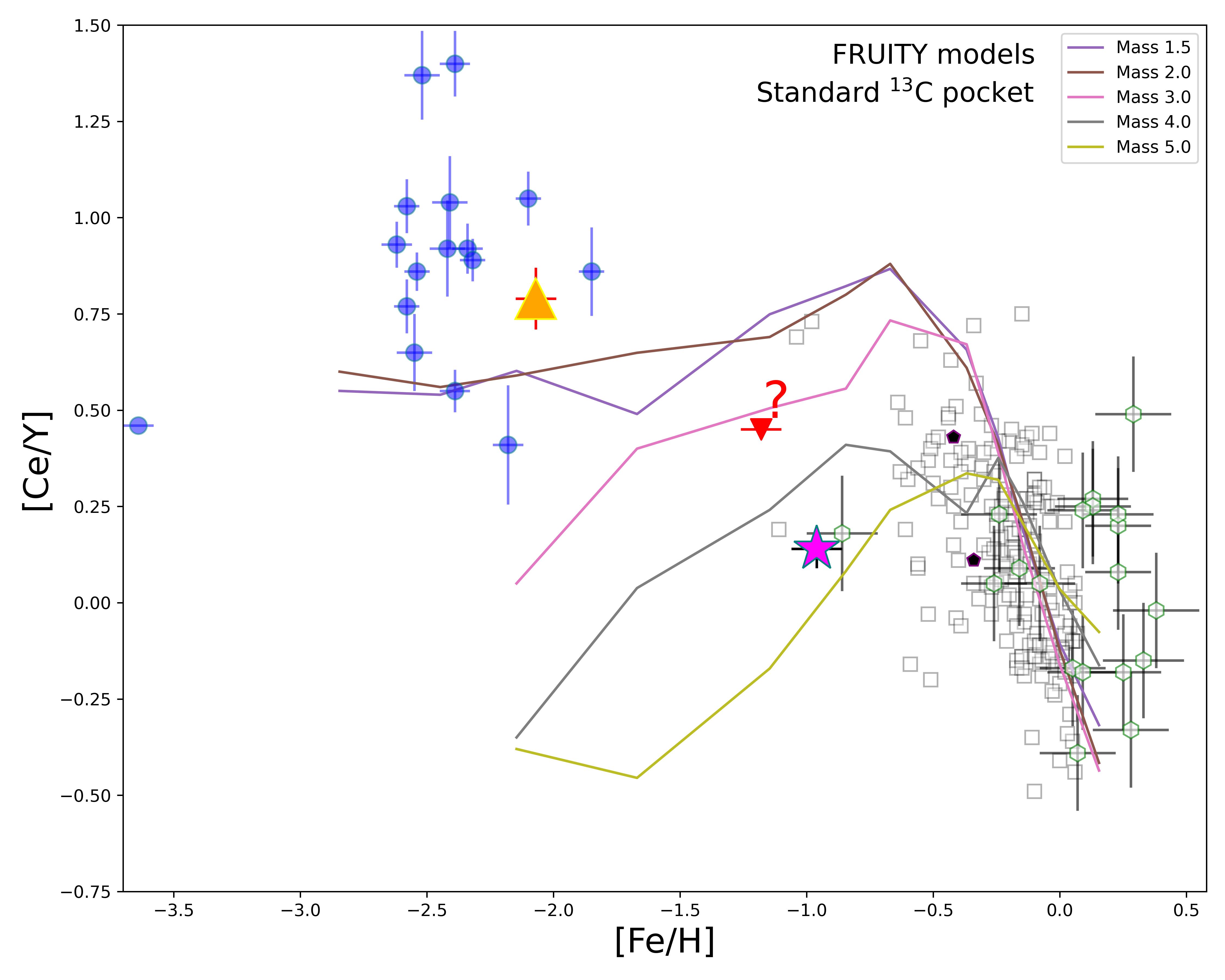}
   \includegraphics[width=0.5\textwidth]{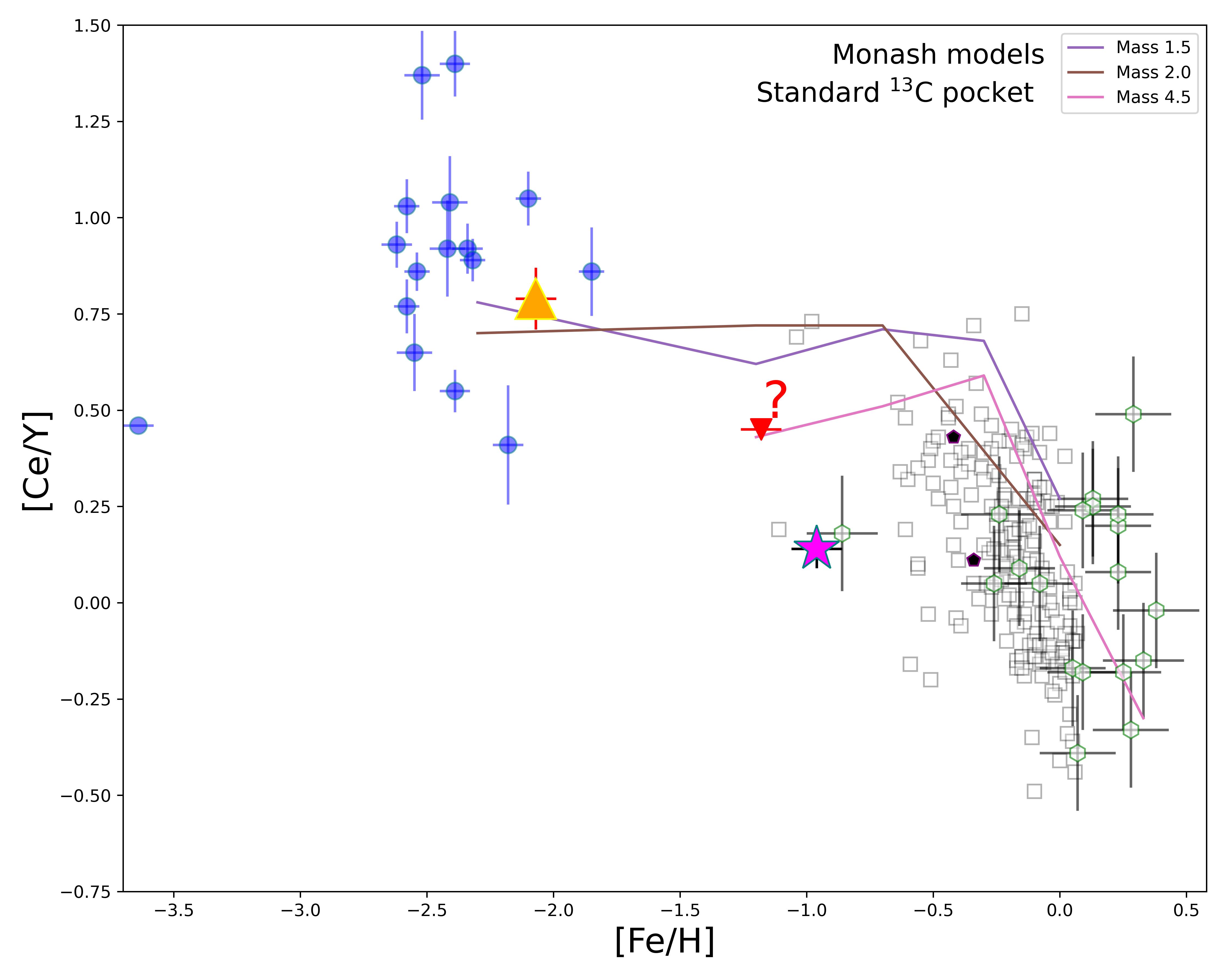}
      \caption{Comparison of observations with stellar models by \texttt{FRUITY} \citep{cristallo2011} and Monash groups \citep[and references therein]{karakas2018}. Symbols as for Figure \ref{fig:cey_feh}, masses are in \msun.}\label{fig:models}
   \end{figure}

In Figure \ref{fig:models}, we present a comparison of the [Ce/Y] ratios as a function of metallicity for our RR Lyrae stars, as well as Ba and CEMP-s stars, alongside AGB models from the \texttt{FRUITY} and Monash groups. 
The \texttt{FRUITY} models \citep{cristallo2009, cristallo2011, cristallo2015, cristallo2016} considered in this study were obtained from their database\footnote{\url{http://fruity.oa-abruzzo.inaf.it/}} using the standard $^{13}$C pocket and not including rotation. The Monash models are based on previous studies by \citet{lugaro2012,fishlock2014,karakas2016,karakas2018}.
We refer the reader to \citet{kappeler2011}, \citet{karakas2014}, and \citet{lugaro2023} for a more detailed discussion on the models and nucleosynthesis of AGB stars. 

In our analysis, we focused on AGB models with masses ranging from 1.5 to 5~M$_\odot$, ensuring they meet the criteria of including third dredge-up episodes while avoiding efficient hot bottom burning (HBB), which would destroy carbon. In the low-mass models (below $\approx$ 4~M$_\odot$, depending on the metallicity) the $^{13}$C($\alpha$,n)$^{16}$O reaction is activated and dominates the neutron production for the s-process in a so-called $^{13}$C pocket. We note that the larger mass (5~M$_\odot$ and above) \texttt{FRUITY} models at low metallicities experience hot bottom burning and hot-third dredge up simultaneously, which prevent the formation of $^{13}$C neutron source \citep{cristallo2015}. 

%dominates as a neutron source producer for the s-process in a so called $^{13}$C pocket. At higher masses, reaching higher temperatures (above 300 \times 10$^6$ K), the $^{22}$N($\alpha$,n)$^{25}$Mg reaction starts to dominate the neutron release.
The previously published Monash grid lacked data for metallicity [Fe/H] = $-$1.2, including a $^{13}$C pocket above 3~M$_\odot$. 
Consequently, we calculated new models at Z = 0.001 ([Fe/H] = $-$1.2) for 4.5~M$_\odot$. %with a partial mixing zone of 1 $\times$ 10$^{-4}$~M$_\odot$, which leads to the formation of the $^{13}$C pocket. 
%The 4.5~M$_\odot$, Z = 0.001 AGB model used in this study was calculated 
using the same methods and procedures outlined in \citet{karakas2016}, where we evolve the 4.5~M$_\odot$ model from the main sequence to the end of the thermally-pulsing AGB phase. We note that the 4.5~M$_\odot$ does have some HBB, but it is relatively mild and the model remains carbon rich. For the evolutionary sequences, we used the same mass-loss rate on the AGB as described in \citet{karakas2018}, which assumes \citet{bloecker1995} with $\eta = 0.01$, including the same input physics as described in that paper. The AGB model experienced 30 thermal pulses, where 29 had efficient third dredge-up. Note that the model of the same mass and metallicity from \citet{fishlock2014} experienced 79 thermal pulses, owing to the use of \citet{VassiliadisWood1993} mass-loss on the AGB instead. Our new model also experienced hot bottom burning, with a peak temperature of around 80 million K.  Post-processing nucleosynthesis calculations were performed to obtain the s-process element abundance predictions, where we include a $^{13}$C pocket via the same method as outlined in \citet{karakas2016}, where the mass of the partially mixing zone of protons was set at $M_{\rm pmz} = 1 \times 10^{-4}$~M$_\odot$. 
In the further discussion, we use the new 4.5~M$_\odot$ Monash model instead of the 4 and 5~M$_\odot$ models without a $^{13}$C pocket. This allows us to compare the observational data only to Monash models that were calculated with the inclusion of a $^{13}$C pocket along the metallicity range of the stars of interest in this study.
%to compare to the \texttt{FRUITY} 4 and 5~M$_\odot$ models and the observational data we use only the 4.5~M$_\odot$ Monash model, since this mass has the most complete set of models with the inclusion of a $^{13}$C pocket along the metallicity range of the stars of interest in this study.

Comparing the different Monash and \texttt{FRUITY} models across the metallicity range, the models generally reproduce the observations well at higher metallicities ([Fe/H] $\gtrsim$ $-$0.5), where most Ba stars are located. In particular, they capture the expected and observationally confirmed declining trend of [Ce/Y] ratios with increasing metallicity, which has been extensively discussed in previous studies \citep{cseh2018,cseh2022,roriz2021}. Overall, the models perform adequately at high metallicity. However, neither set of models can explain the elevated [Ce/Y] ratios observed in CEMP-s and TY Gruis stars. At metallicities below $-$2, the models underestimate the ratios for any given AGB mass. Although a comprehensive examination of the different model sets is beyond the scope of this paper, we also checked the rotating models. We found that they do not adequately explain the patterns observed at low metallicities. Further investigations will be necessary.

As shown in Figure \ref{fig:models}, our two RR Lyrae stars are positioned in the intermediate range of the metallicity distribution, i.e., [Fe/H] $\approx -1$. 
Based on the model predictions, the {\texttt FRUITY} grids for masses between 1.5 and 4 \msun~can reproduce the observed distribution, once the dilution of AGB material is taken into account. Similarly, Monash models for 1.5, 2.0, and 4.5 \msun~can all explain the observed abundance patterns in the [Ce/Y] versus [Fe/H] plane, again assuming that dilution effects are considered (see following discussion). Following \cite{cseh2022}, we have calculated the dilution factors (which take into account the mass transfer and the mixing) of AGB abundances using the relationship:
\begin{equation}
{\rm [X/Fe]_{\rm dil} = log(10^{[X/Fe]_{\rm ini}} \times (1 - \delta) + 10^{[X/Fe]_{\rm AGB}} \times \delta)}, 
\end{equation}
where ${\rm [X/Fe]_{ini}}$ represents the initial abundances in the models, assumed to be solar for all species; ${\rm [X/Fe]_{\rm AGB}}$ gives the final abundances produced by the AGB; and ${\rm [X/Fe]_{\rm dil}}$ is the abundance measured in our RRLs DQ Hya and TY Gruis. We omit ASAS J153830-6906.4 in this context due to uncertainties related to its barium abundance, which we use as a proxy for Ce; this element was not measured in the HERMES spectra.

We then calculated the $\delta$ value, defined as $1/\text{dil}$, 
which represents the ratio between the total mass, i.e., the original envelope mass of the Ba star plus the mass transferred from the
AGB star to its companion, and the transferred AGB material,
and thus ${\rm dil= M_{total}/M_{AGB,trans} = (M_{Ba,env} + M_{AGB,trans})/M_{AGB,trans}}$ (see \citealp{cseh2022} and references therein). Thus, a value of $\delta$=0 corresponds to no mass transfer from the AGB star, and a value of $\delta$=1 represents the unrealistic case of no mixing of the AGB material into the RRL star after the mass transfer. Typical values reported in \cite{cseh2022} and \cite{yang2024} are $\delta=0.3$; however, in some cases, larger values can be derived (see discussion in the corresponding papers). 
We constrained the dilution factors using both [Ce/Fe] and [Y/Fe] abundances, with the expectation that they should yield consistent $\delta$ values. However, this is not the case for most of our calculations, so caution is advised when interpreting these dilution factors. In particular, for the metal-poor TY Gruis, the [Y/Fe] ratio results in $\delta$ values typically ranging from 0.02 to 0.1 for both Monash and \texttt{FRUITY} models across nearly all stellar masses. This is expected, given the low metallicity of the star, where even a small amount of transferred mass can significantly alter the surface abundances of the companion. 
However, the \texttt{FRUITY} models do not produce sufficient Ce (as previously noted and clearly shown in Figure \ref{fig:models}); therefore, when calculating the same dilution factor using [Ce/Fe] abundances, we obtain $\delta$ values for masses of 1.5 and 2 M$_\odot$ that are a factor of 3 higher than those derived from [Y/Fe], while yielding unrealistic dilution values for masses of 3 M$_\odot$ and above.
Conversely, the Monash models tend to converge toward similar $\delta$ values for masses between 1.5 and 3 M$_\odot$; for masses between 4 and 5 M$_\odot$, the production of Ce becomes very limited and unphysical $\delta$ values of larger than 0.7 would be required. For DQ Hya, the \texttt{FRUITY} models would require $\delta$ values greater than 1 across all masses for Y, while [Ce/Fe] abundances would require $\delta$ values of 0.48, 0.25, and 0.58 for 1.5, 2.0, and 3.0 M$_\odot$, respectively. In contrast, the Monash models suggest $\delta$ of 0.82, 0.37, and 0.86 for the [Y/Fe] ratios at masses of 1.5, 2.0, and 3.0 M$_\odot$, respectively. However, for [Ce/Fe] ratios, the $\delta$ values are estimated at 0.26, 0.09, and 0.24 for 1.5, 2.0, and 3.0 M$_\odot$, while higher masses imply unrealistically large $\delta$ values exceeding 1, indicating these more massive AGB stars do not produce enough Ce, consistent with a decreasing TDU efficiency and smaller $^{13}$C pocket.

Due to the substantial discrepancies observed in the calculated dilution factors, we are unable to definitively determine the mass of the AGB companion at this time; additional studies are required. Still, some limits on the amount of accreted material can be inferred by noting that DQ Hya and the other two s-rich RRLs will pass through the instability strip while helium is burning in their core, as RR Lyrae stars. Its pulsation and evolutionary characteristics can thus provide constraints on the possible accreted mass. We discuss this in more detail in the following Section \ref{sec:formation}.

\subsection{Formation scenarios}\label{sec:formation}

For many years, RR Lyrae stars were considered reliable tracers of old stellar populations due to their predominant presence in the Galactic halo, globular clusters, and generally in ancient stellar systems. However, this paradigm has been challenged by the discovery of metal-rich RR Lyrae stars exhibiting thin disc kinematics and low $\alpha$-element abundances \citep[and references therein]{layden1995, chadid2017,prudil2020, crestani2021b,iorio2021,gozha2021,gozha2024,dorazi2024}. There is an ongoing debate in the literature regarding the formation channels of metal-rich RR Lyrae stars (see summary in \citealt{zhang2025}). The current hypotheses include single-star evolutionary pathways, possibly involving extragalactic origins to account for their peculiar chemical signatures \citep{gozha2021,gozha2024,dorazi2024}, and binary formation scenarios \citep{bobrick2024}. In the latter case, material transfer may occur via mechanisms such as accretion through a circumbinary disc. Our two RR Lyrae stars exhibiting s-process enrichment, occupy a parameter space in metallicity ([Fe/H] $\approx -1$) where single-stellar evolutionary models typically predict their existence. Nevertheless, several intriguing hypotheses warrant further consideration. Comparisons with AGB models suggest that the RR Lyrae progenitors may have accreted material from an approximately $\approx$1.5–5.0 \msun\, AGB companion. Considering the rapid evolutionary timescales of more massive stars, this indicates that mass transfer likely occurred while the secondary star (the current RR Lyrae) was still on the main sequence. 

Comparing the luminosity (Section \ref{sec:puls}), the spectroscopic temperature, and the log g measurements (Table \ref{tab:hermes_abundances}) with PARSEC tracks \citep{nguyen2022}, the star ASAS J153830-6906.4 is consistent with an HB star of about 0.65 M$\odot$, located close to the ZAHB. A similar result is obtained for DQ Hya if its luminosity is assumed to follow the $M_G$–[Fe/H] relation for field RR Lyrae stars \citep{garofalo2022}.
In contrast, adopting the higher luminosity inferred from the parallax-based distance ($\log L/\mathrm{L}\odot > 1.8$, see Section \ref{sec:puls}), DQ Hya is instead consistent with a late evolutionary phase of an HB star of about 0.58–0.60 M$_\odot$. Assuming a typical RGB mass loss of about 0.20–0.25 M$\odot$ at the metallicity of the RR Lyrae \citep{gratton2010,tailo2021}, the progenitor mass at the onset of the red giant branch would be in the range 0.8–0.9 M$\odot$.
Under this scenario, assuming a typical $\delta$ of 0.3 \citep{cseh2022}, which translates to a dilution factor of roughly 3, approximately 0.1 \msun\ would have been accreted. This aligns with standard accretion models depicted for Ba giant stars. However, due to the significant uncertainty in dilution factors, we avoid making any definitive conclusions at this point.
Moreover, the scenario assumes that after the initial pollution event, no further binary interactions occurred.

Another plausible scenario involves a second episode of mass transfer occurring as the RR Lyrae progenitor ascends the red giant branch for the first time. In this case, the resulting RR Lyrae could be classified as a binary-produced RR Lyrae, as discussed by \cite{bobrick2024}, but with the key difference that it experienced two distinct mass transfer events. The first occurred when the star was on the main sequence, accreting material from an AGB companion and producing the observed chemical anomalies. The second transfer happened later during its RGB phase, where the star, now a giant, transferred some of the previously accreted material back, potentially to the current white dwarf companion. If this scenario holds, the second mass transfer could impose constraints on the current properties of the binary system. Following \cite{bobrick2024}, for the RR Lyrae progenitor to interact with its white dwarf companion, the initial orbital period would need to be shorter than approximately 700 days. After this interaction, the system’s orbital period would likely be on the order of 1,000 days.
In addition, within this scenario the mass lost during the RGB is significantly larger ($\gtrapprox0.4$ M$_\odot$) than expected from single stellar evolution, implying more massive progenitors for the RR Lyrae stars.

An additional scenario is that the current RR\,Lyrae are a member of the sub-luminous post-AGB/post-RGB family crossing the instability strip \citep[see e.g.][]{oomen2018}. Indeed, using Gaia DR3, \citet{Oudmaijer2022} showed that a subgroup of stars previously classified as post-AGBs were too sub-luminous for the expected post-AGB evolution ($\gtrsim 10^2\,L_\odot$) and more likely post-RGBs, descendants of RGB interactions. Without speculating about the origin of these sub-luminous post-AGB-like systems, we note that the current RR Lyrae have surface temperatures, luminosities and abundances indeed consistent with this population, thus suggesting a shared origin. Partially supporting this interpretation is that DQ Hya seems to have a slightly higher absolute magnitude than other field RR Lyrae with a similar metallicity (see Section \ref{sec:puls}), although the ASAS star is consistent in temperature and luminosity with the regular RR Lyrae in globular cluster M4. It may also be worth noting that one of the few RRLs confirmed to be in a binary system is a post-RGB star, a star whose envelope was stripped during the RGB phase and is now crossing the instability strip \citep{BEP}. We also note that several other sub-luminous post-AGB-like stars from \citet{Oudmaijer2022} are also located in the luminosity--temperature region surrounding the RRL instability strip ($1 < \log L/L_\odot < 2$, $3.7 < \log T/\mathrm{K} < 3.9$). We compared the Hertzsprung-Russell diagram positions of these stars with the theoretical RR Lyrae IS boundaries computed by \citet{marconi2015}, and then searched for light curves among the observations of the TESS (Transiting Exoplanet Survey Satellite, \citealt{Ricker-2015}) mission. TESS can detect RR Lyrae stars over a wide brightness range and even in crowded fields with high fidelity \citep{Molnar-2022}. Four out of the five stars within or near the IS boundaries do not show any large-amplitude pulsations that would resemble an RR Lyrae star. The fifth target has not been observed by TESS yet, but it was not detected as a variable by the Gaia mission, strongly suggesting that it is not pulsating as an RR Lyrae, either \citep{GaiaDR3-2023,clementini2023}. Additionally, observations of post-AGB stars are often associated with infrared excesses resulting from the recently stripped material \citep[see e.g.][]{dellagli2023}. However, neither of the two RR Lyrae stars examined in this work exhibits any signs of such excess (see the WISE photometry for DQ Hya shown in Figure \ref{fig:sed_dqhya}). Therefore, the association of the current RR Lyrae with this population should be examined carefully before drawing any conclusions.

Overall, none of the scenarios discussed above can be confirmed as the definitive formation pathway for these two RR Lyrae stars. In particular, if DQ Hya is truly an evolved RR Lyrae, this raises an important point to consider. The time spent by an HB star of initial mass 0.58 M$_\odot$ in the instability strip is more than two orders of magnitude shorter than that of a canonical RR Lyrae at the same metallicity. Simple statistical reasoning therefore implies that it is already rare to find such an object in a sample of $\approx 500$ RR Lyrae analysed in this work. It is even more unlikely to encounter it within the subsample of only two stars showing peculiar abundances, unless the high luminosity of DQ Hya and its chemical peculiarity are connected through a distinct formation channel.

Future works will need detailed binary evolution modelling, specifically tailored to the metallicity, masses, and orbital parameters relevant to these systems. This will include considerations of multiple mass transfer episodes, angular momentum loss, and surface abundance evolution. Such work will be presented in a dedicated upcoming paper.

\subsection{Search for binary signatures}\label{sec:astro}

Since Ba stars are known to be members of binary systems, their orbital parameters can often be constrained through RV observations \citep{jorissen2019, escorza2019}. The RV variations over time provide direct evidence of binarity and allow for the determination of orbital periods, eccentricities, and mass functions. However, for RR Lyrae stars, measuring RVs is significantly more challenging. This difficulty arises from their large-amplitude pulsations, which cause substantial and periodic changes in their spectral lines that can complicate the detection of orbital-induced RV variations. Consequently, traditional spectroscopic methods are less effective for these variables. As an alternative approach, we can utilise astrometric data from \textit{Gaia} to search for signatures of binarity. \textit{Gaia}'s precise measurements of parallaxes and proper motions can reveal orbital motions on the sky, even when RV data are unavailable or inconclusive. 

Additionally, the coherent pulsations of RR Lyrae stars themselves can be exploited to infer orbital parameters from photometry alone. This method involves analysing the light-time effect -- the slight shifts in pulsation arrival times caused by the star’s orbital motion around a common centre of mass. By carefully monitoring the periodic variations in pulsation timings, it is possible to detect the presence of a companion and estimate orbital elements, even without spectroscopic data.

Unfortunately for RR Lyrae stars, the Blazhko effect, the quasi-periodic modulation of the pulsation amplitude and phase, can mask any O--C signal caused by the light-time effect \citep{Benko-2014}. Furthermore, many RR~Lyrae stars show O--C curves that are characterised by largely unexplained rapid changes and sudden breaks, although others show clear secular period changes \citep[see, e.g.,][]{leborgne-2007}. However, the feasibility of the light-time method for non-Blazhko RR Lyrae stars was proven by \citet{hajdu2015,hajdu2021}, who identified several RR Lyrae binary candidates. O--C curves of the stars can thus reveal both the presence and distance of a companion and/or the rate and direction of the variable's evolution across the instability strip, thereby helping to constrain different evolutionary scenarios. This work will be presented in a dedicated publication. 

\subsection{Astrometric signature}

   \begin{figure}
   \centering
   \includegraphics[width=1.0\columnwidth]{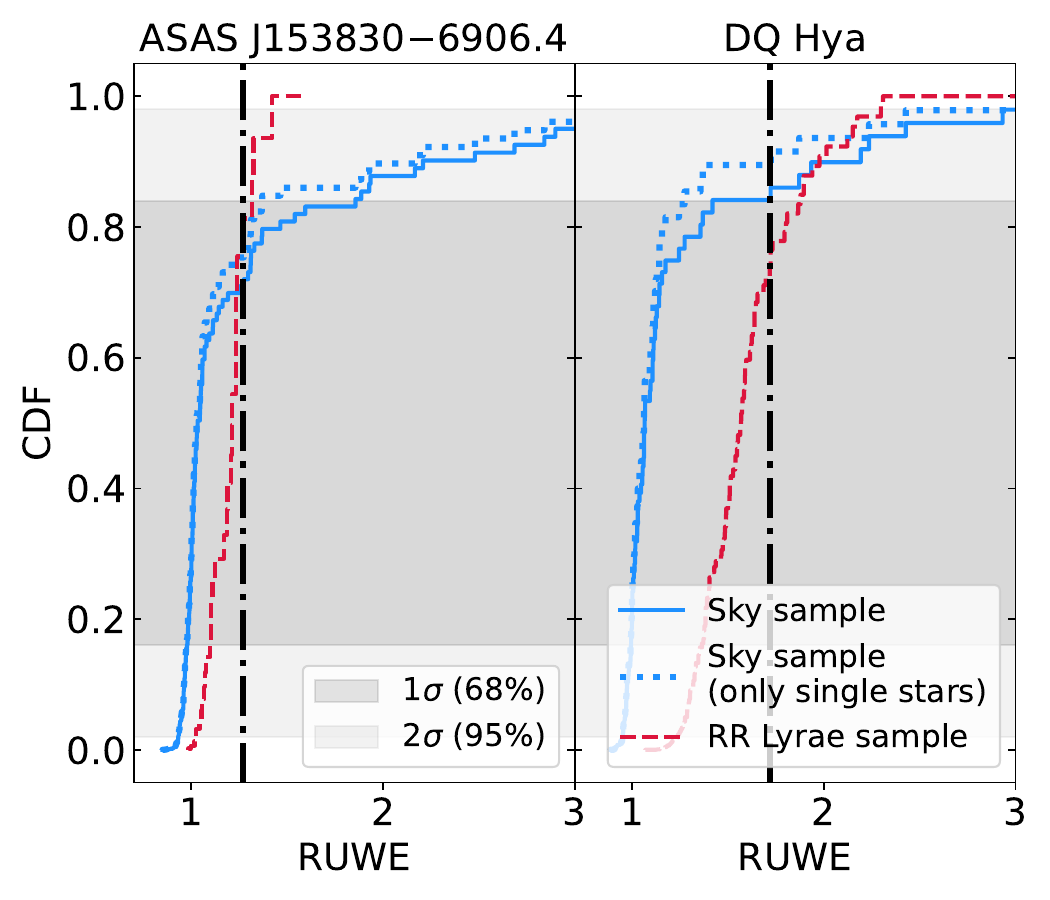}
    \caption{RUWE values (vertical black dash--dot lines) of the two analysed RR\,Lyrae stars (left: ASAS\,J153830-6906.4, right: DQ\,Hya) compared to the cumulative distribution functions (CDFs) of the corresponding control samples. 
    The blue solid line shows the CDF for the sky-position control sample, which includes stars in the same sky area with similar photometry. 
    The blue dotted line shows the CDF for the same sample after removing stars classified as non-single in Gaia~DR3. 
    The red dashed line shows the CDF for the RR\,Lyrae control sample from Gaia~DR3, selected to have similar photometry and light-curve properties to the targets (see text for details). 
    The shaded regions indicate the central 68\% (dark shading) and 95\% (light shading) intervals of the distributions.}
    \label{fig:ruwe}
   \end{figure}

Unresolved binary systems can be detected through precise astrometric measurements, as the orbital motion of the photocenter causes systematic deviations from the behaviour of a single star. Gaia DR3 has provided the first catalogue of astrometric binaries \citep{halbwachs2023,holl2023}. However, neither of the two RRLs analysed in this work is flagged as an astrometric binary in Gaia DR3. However, this does not rule out the presence of companions, as Gaia’s astrometric binary detection is not sensitive to all orbital configurations \citep{elbadry2024}.
In addition, in Gaia DR3 no binaries solution are reported for binary periods  $P_\mathrm{orb} < 20000 \ \delta \varpi/\varpi $ days ($\varpi/\delta \varpi$ is the parallax over its error),  corresponding to $P_\mathrm{orb}\lessapprox 1700$ days for DQ Hya ($\varpi/\delta \varpi\approx12$) and $P_\mathrm{orb}\lessapprox 800$ days for ASAS J153839-6906.4 ($\varpi/\delta \varpi\approx26$). 
Moreover, Gaia DR3 has detected very few astrometric binaries beyond roughly 2 kpc, and since both stars in this study are located at distances greater than that, their intrinsic likelihood of being detected as binaries by Gaia’s current pipeline is relatively low.

Another commonly used diagnostic is the Renormalised Unit Weight Error (RUWE), which provides a normalised assessment of how well Gaia’s astrometric data fit a single-star model (similar in concept to a reduced $\chi^2$). Solutions for well-behaved, single sources typically have RUWE values close to 1, while higher values suggest a poorer fit to the single-star assumption. Elevated RUWE can result from crowding, background contamination, or intrinsic variability. One of the main causes is unmodelled binarity, where the orbital motion of the photocenter is not properly accounted for \citep{belokurov2020, penoyre2022, castro2024}.
A RUWE threshold of 1.4 is often adopted to distinguish between reliable and potentially biased astrometric solutions \citep{halbwachs2023}. For our targets, we find RUWE=1.27 for ASAS J153839-6906.4 and RUWE=1.72 for DQ Hya. The first value lies within the typical range for single stars, while the second exceeds the 1.4 threshold, suggesting a possible bias in the astrometric fit and raising the possibility of an unresolved companion.
DQ Hya also shows a significant difference between the distance estimate with parallax and that from the magnitude-metallicity relation (see Section \ref{sec:puls}), supporting the scenario in which the parallax estimate is biased by the presence of a companion.  However, the significance of the RUWE can be affected by photometric properties and sky position \citep{castro2024}.  Variability can also influence the RUWE distribution: \citet{belokurov2020} found that, for the Gaia~DR2 RR\,Lyrae sample, the median RUWE depends systematically on pulsation properties and apparent magnitude.  
Therefore, to assess the significance of the measured RUWE values for our targets, we compare them against two control samples:

\begin{itemize}
    \item Sky control sample: All Gaia DR3 sources within 1º of the analysed RR Lyrae stars, with G-band magnitude within 0.5 mag and Gaia colour within 0.1 mag of the target. We also define a subset of the control sample in which we remove all the stars classified as astrometric non-single stars in Gaia DR3.

    \item RR Lyrae control sample. All Gaia DR3 RR Lyrae variables with light-curve properties similar to the two analysed RR Lyrae stars: periods within 0.15 days, G-band peak-to-peak amplitudes within 0.15 mag, $\phi_{31}$ light-curve parameter angles\footnote{The $\phi_{31}$ is a Fourier phase parameter derived from the decomposition of the light curve, see \cite{clementini2023}.} within 0.2 rad, and the same magnitude and colour cuts as for the sky control sample.

\end{itemize}

Figure~\ref{fig:ruwe} shows the comparison between the RUWE values of the two analysed RR\,Lyrae stars and those of the control samples.  
For ASAS\,J153830$-$6906.4, the RUWE lies well within the typical distribution for both the Sky and RR\,Lyrae control samples.  
In contrast, DQ\,Hya is located closer to the tail of the distribution for the Sky control sample, particularly when considering the version cleaned of sources classified as non-single in Gaia~DR3.  
However, even in this case, the RUWE value remains within the 95\% confidence interval of the distribution.  
The comparison with the RR\,Lyrae control sample instead shows that the DQ\,Hya RUWE is typical of RR\,Lyrae stars with similar light-curve properties, i.e. with short periods (\(\approx 0.5\)~days) and large amplitudes (\(\gtrapprox 1\)~mag in the \(G\)~band), which tend to exhibit higher RUWE values, as discussed in \citet{belokurov2020}.  

In conclusion, while the RUWE values are somewhat higher than average, they are not indicative of a significant bias in the astrometric fit that would suggest the presence of a binary.  
This alone cannot exclude an undetected binary companion, but it does help to constrain the possible orbital period: the sensitivity of RUWE to binarity peaks for periods of $\sim 3$~years and decreases rapidly, becoming negligible for periods longer than $\sim 30-50$~years \citep{penoyre2022,castro2024}.  

An additional consequence of this analysis is that we can consider the Gaia~DR3 astrometric solutions to be robust for both stars.  
In particular, this supports the conclusion that DQ\,Hya is intrinsically brighter than typical RR\,Lyrae stars of similar metallicity in the field (see Section~\ref{sec:puls}).

\section{Concluding remarks}

In this work, we have discovered and analysed two new RR Lyrae stars exhibiting significant s-process element enrichment, expanding the rare class of such objects (2 in 470 of our sample) beyond the previously known TY Gruis \citep{preston2006}. Based on detailed elemental abundance analyses, we suggest that these stars may have experienced mass transfer from former AGB companions, supporting their classification as post-mass transfer systems. However, several uncertainties persist. The precise properties of the putative AGB companions, such as their mass and the extent of accreted material, are still unconstrained by model limitations and uncertainties in dilution factors. Different formation scenarios, including mass transfer during the main sequence, second mass transfer episodes on the red giant branch, or the possibility of a sub-luminous post-AGB-like crossing the instability strip, offer plausible explanations but lack definitive confirmation. Our investigation highlights the importance of combining chemical, dynamical, and astrometric data to understand these complex objects. 

Although Gaia's current data do not provide conclusive evidence of binarity for these stars, partly due to observational limitations at their distances, their RUWE values and the consistency with chemodynamical expectations suggest that undetected companions cannot be entirely ruled out. Pulsations can also be used to detect binarity. We searched for time series photometry of both targets in various public databases and in the observations of the TESS (Transiting Exoplanets Survey Satellite, \citealt{Ricker-2015}) mission. Unfortunately, ASAS J153830-6906.4 turned out to be a known Blazhko star \citep{szczygiel-2007,skarka-2013}. DQ~Hya, however, appears to be a pure fundamental-mode pulsator based on its TESS light curves, and thus amenable to such analysis. A detailed investigation of the O--C curve will be published separately.

Future dedicated binary evolution modelling, along with continued monitoring of orbital and pulsation timing variations, will be essential to elucidate the formation pathways of these intriguing RR Lyrae stars. These efforts will be addressed in upcoming work, aiming to deepen our understanding of stellar evolution and binary interactions of variable stars.

\begin{figure}
    \centering
    \includegraphics[width=1.0\linewidth]{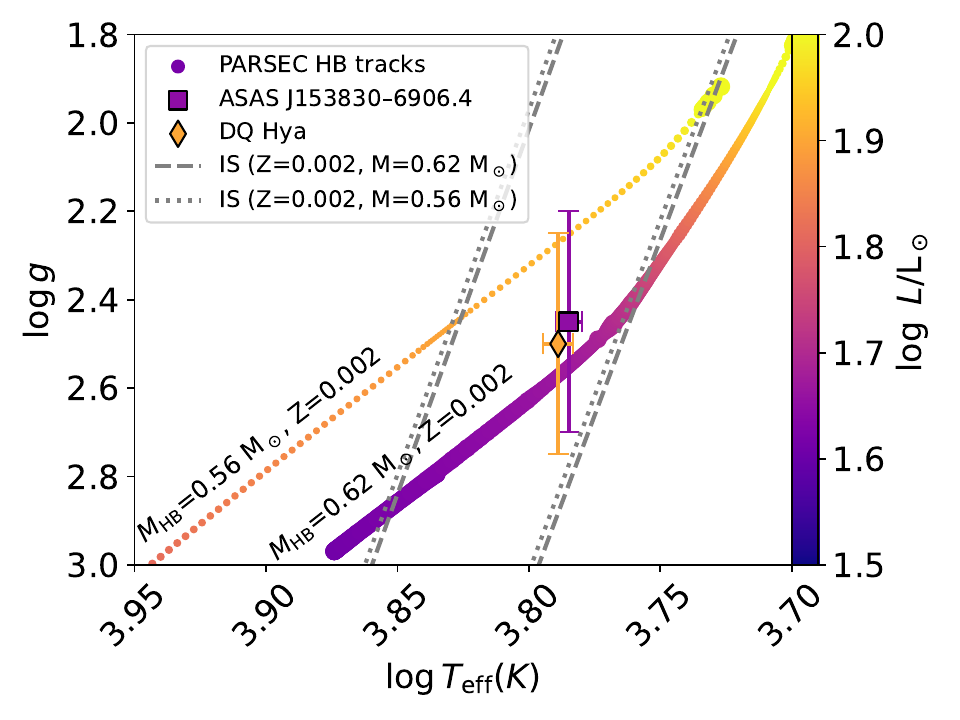}
    \caption{Kiel diagram comparing the RR Lyrae stars analysed in this work (DQ Hya shown as a diamond, ASAS J153830-6906.4 as a square) with horizontal branch evolutionary tracks from the PARSEC V1.2S database  (\citealt{chen2015}, circles; masses and metallicities are indicated in the plot). The colormap represents the bolometric luminosity. For the observed stars, luminosities are derived using distances from parallaxes (see Section~\ref{sec:puls}). The size of the circles along the tracks provides a qualitative indication of the time spent in each region of the diagram, with the largest symbols corresponding to approximately ten times longer evolutionary timescales than the smallest ones. The grey lines indicate the instability strip from \citet{marconi2015}, assuming a metallicity of $Z = 0.002$ and stellar masses of $0.56,M_\odot$ (dotted line) and $0.62,M_\odot$ (dashed line). The comparison is purely qualitative and does not represent the result of a fit.
}
    \label{fig:HD}
\end{figure}

\begin{acknowledgements}
    We thank the reviewer for their valuable comments and suggestions that improved the quality of this manuscript. GI thanks Pau Ramos for fruitful discussions on the kinematic association with Galactic components.
    Part of this work was supported by the Fulbright Visiting Research Scholar program 2024-2025.  GI was supported by a fellowship grant from la Caixa Foundation (ID 100010434). The fellowship code is LCF/BQ/PI24/12040020. This research was supported by the `SeismoLab' KKP-137523 \'Elvonal grant of the Hungarian Research, Development and Innovation Office (NKFIH), by the NKFIH excellence grant TKP2021-NKTA-64 and by the LP2025-14/2025 and LP2023-10 Lendület grants of the Hungarian Academy of Sciences. A.B.~acknowledges support from the Australian Research Council (ARC) Centre of Excellence for Gravitational Wave Discovery (OzGrav), through project number CE230100016. SWC acknowledges federal funding from the Australian Research Council through a Future Fellowship (FT160100046) and Discovery Projects (DP190102431 and DP210101299). This research made use of NASA’s Astrophysics Data System Bibliographic Services, as well as of the SIMBAD and VizieR databases operated at CDS, Strasbourg, France.
\end{acknowledgements}

\bibliographystyle{aa} 
\bibliography{rrl} 

\clearpage

\begin{appendix} %First appendix
\section{Supplementary material}

\subsection{Linelist and atomic data}
\begin{figure}[h!]
    \centering
    \includegraphics[width=0.5\textwidth]{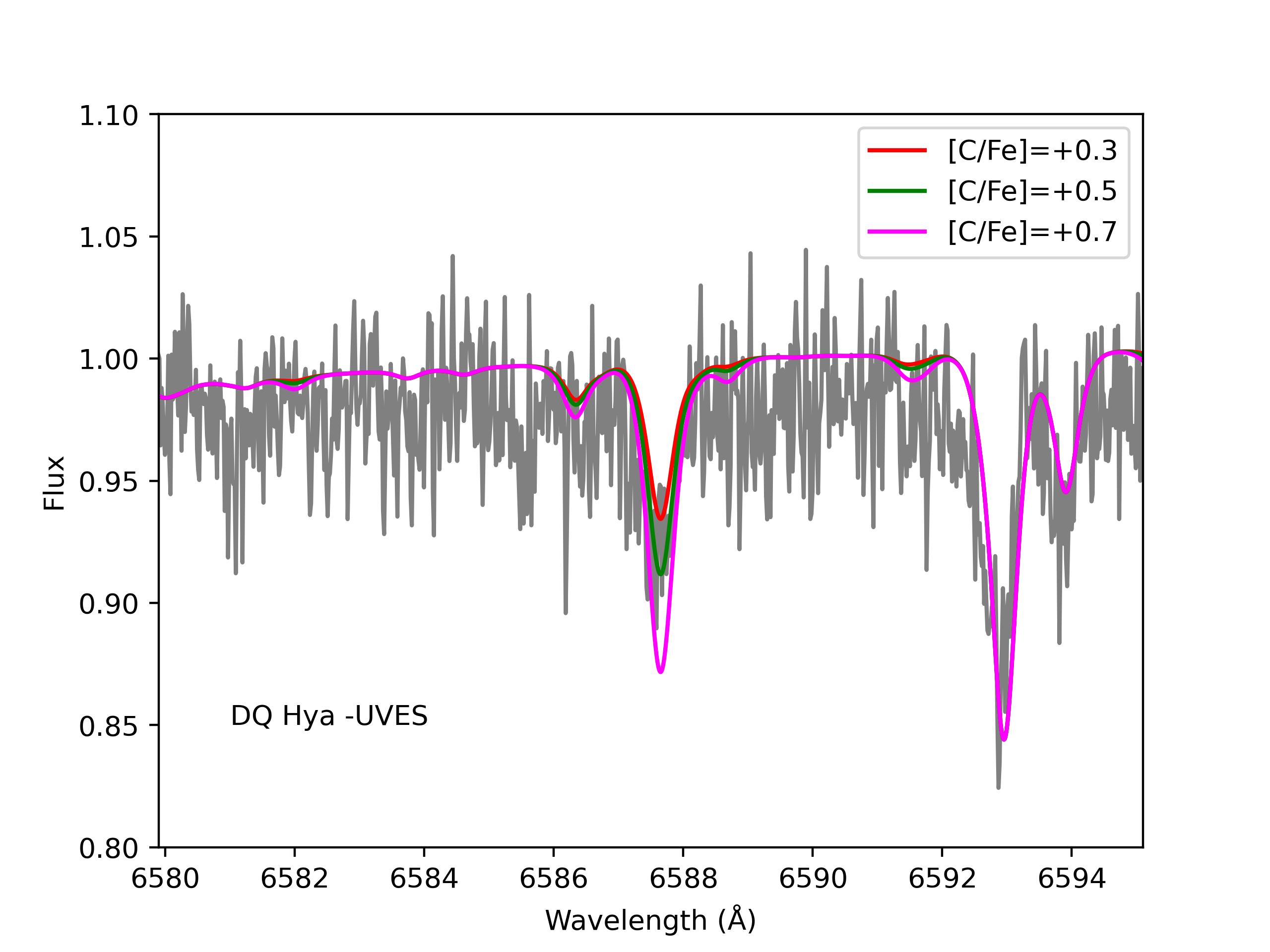}
    \includegraphics[width=0.5\textwidth]{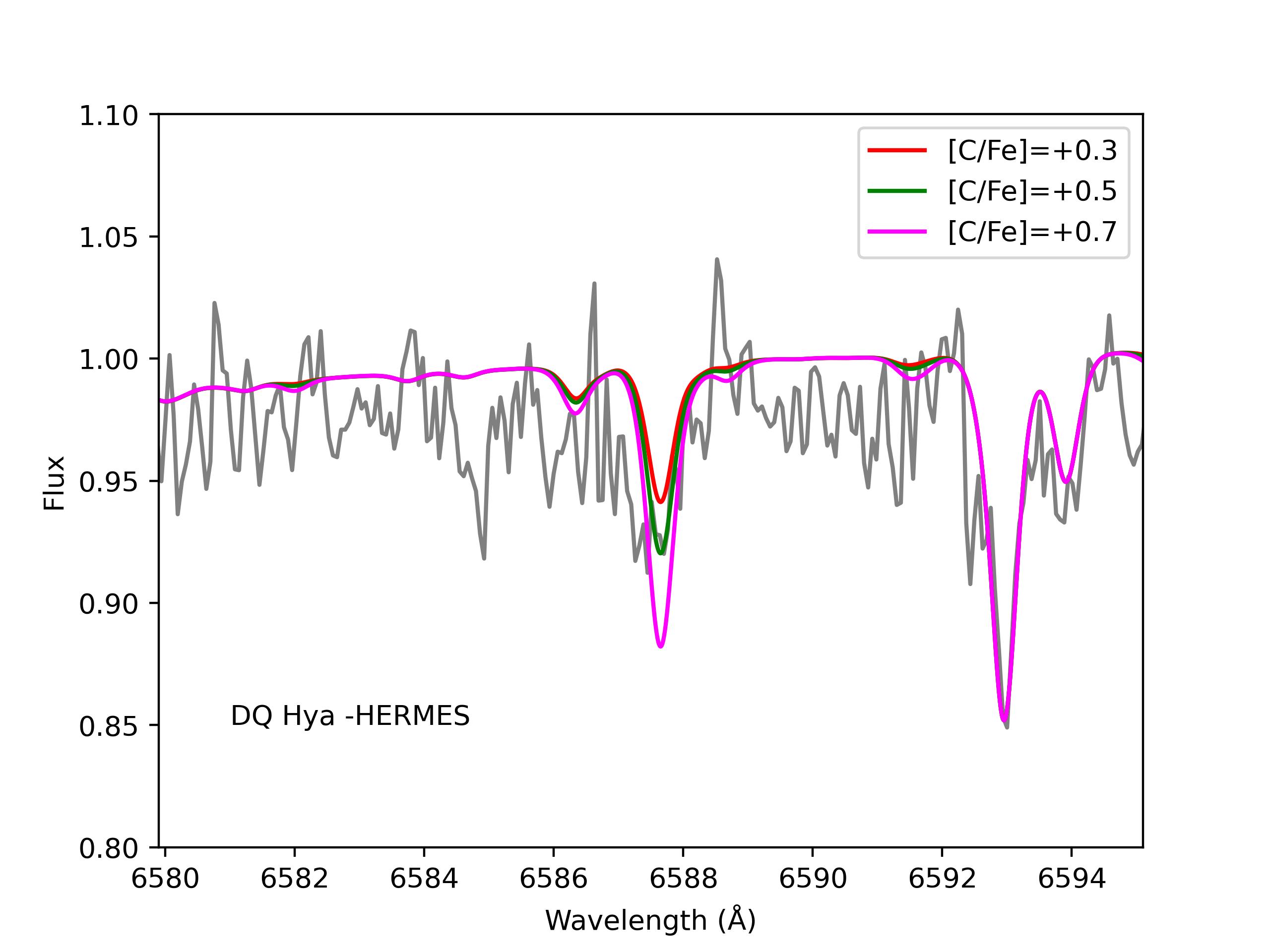}
    \caption{Example of spectral synthesis for C I line at 6587 in DQ Hya for both HERMES and UVES spectra. There is a remarkable agreement between the two estimates. We note that NLTE departures are negligible for this line, especially at the parameter range of our sample stars \citep{alexeeva2015}.}
    \label{fig:synthesis_carbon}
\end{figure}

\begin{figure*}[h!]
    \centering
    \includegraphics[width=0.9\textwidth]{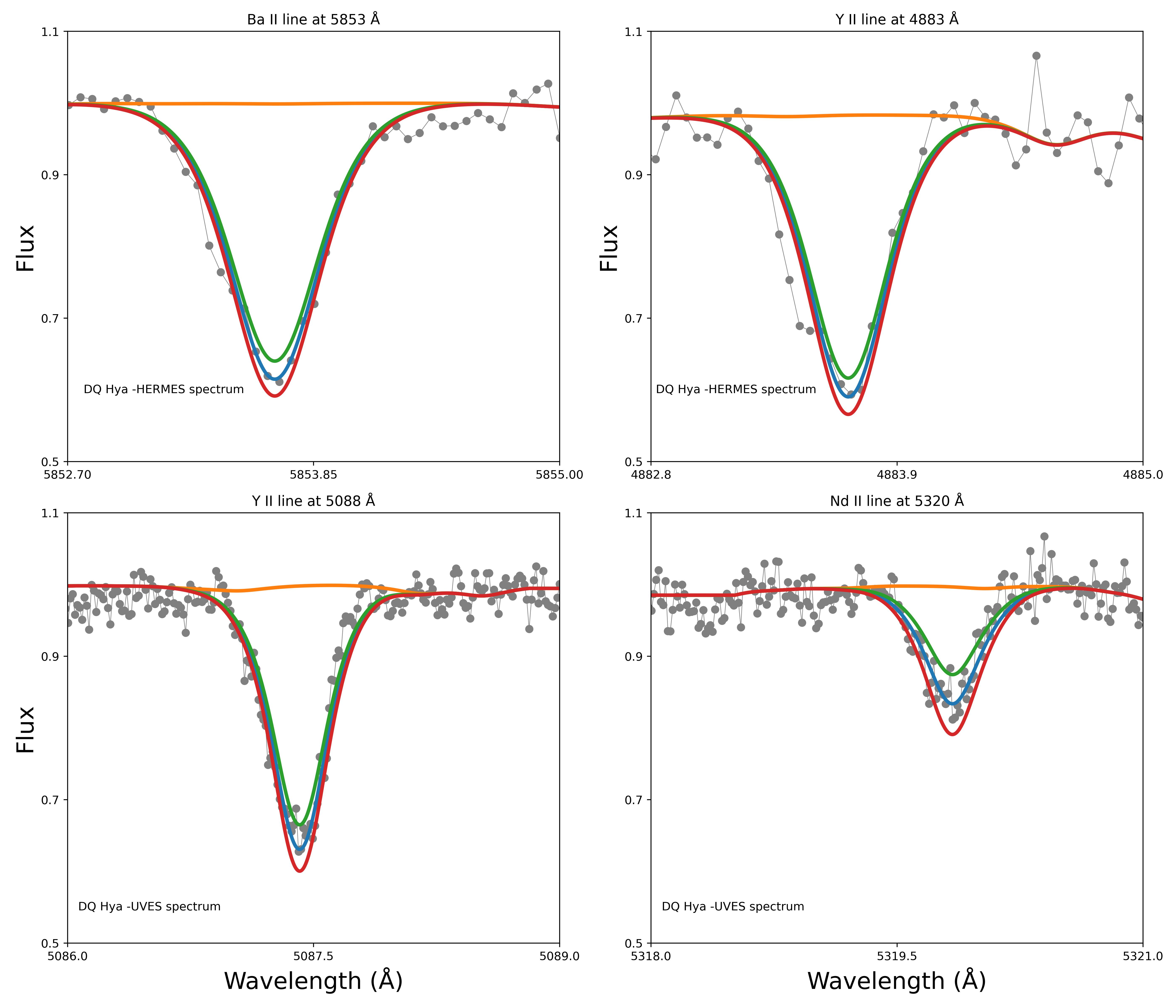}
    \caption{Example of spectral synthesis for Ba, Y, and Nd lines in DQ Hya. The orange lines represent nearly zero abundances for all the species in each panel. The blue lines denotate the best-fit abundances, while red and green lines are for $\pm$ 0.2 dex, respectively.}
    \label{fig:synthesis}
\end{figure*}

In Tables \ref{tab:linelist} and \ref{tab:linelist_uves} we report linelist and atomic data used for parameter and abundance determination from HERMES/GALAH and UVES spectra, respectively.
Each table contains wavelength, species, excitation potential and transition probabilities ($\log gf$). Examples of the spectral fitting procedures are given in Figures \ref{fig:synthesis_carbon} 
and \ref{fig:synthesis}.

\begin{table}[h]
    \centering
    \caption{Line list (and corresponding atomic data) for parameter and abundance determinations in HERMES/GALAH spectra. Hyperfine and isotopic splitting have been included for Ba II lines, for which we report the total oscillator strengths of the atomic transition here.}\label{tab:linelist}
    \begin{tabular}{lccr}
    \hline
    Line & Species & E.P. (eV) & $\log gf$ \\
    \hline
    6587.610 & C {\sc i}  & 8.537  & -1.05\\
    5711.088 & Mg {\sc i}  & 4.346 & -1.742 \\
    5857.451 & Ca {\sc i}  & 2.933 &  0.240 \\
    6493.781 & Ca {\sc i}  & 2.521 & -0.109 \\
    6499.650 & Ca {\sc i}  & 2.523 & -0.818 \\
    4820.409 & Ti {\sc i}  & 1.503 & -0.380 \\
    5866.451 & Ti {\sc i}  &  1.067 & -0.790   \\
    4719.511 & Ti {\sc ii} & 1.569 & -1.72 \\
    4764.525 & Ti {\sc ii} & 1.237 & -2.690  \\
    4798.531 & Ti {\sc ii} & 1.080 & -2.660  \\
    4849.168 & Ti {\sc ii} & 1.131 & -2.960 \\
    4865.610 & Ti {\sc ii} & 1.116 & -2.700 \\
    6513.044 & Ti {\sc ii}  & 4.002 & -1.490 \\
    4727.413 & Fe {\sc i}   & 3.686 & -1.083 \\
    4736.774 & Fe {\sc i}   & 3.211 & -0.674 \\
    5679.023 & Fe {\sc i}   & 4.652 & -0.820 \\
    5705.465 & Fe {\sc i}   & 4.301 & -1.355 \\
    6546.239 & Fe {\sc i}   & 2.759 & -1.536  \\
    6593.870 & Fe {\sc i}   & 2.433 & -2.420  \\
    6516.077 & Fe {\sc ii}  &2.891 & -3.310 \\
    4883.682 & Y {\sc ii}   & 1.084  & 0.190  \\
    5853.668 & Ba {\sc ii}  & 0.604 & -0.907 \\
    6496.897 & Ba {\sc ii} &  0.604 & -0.407  \\
    \hline
    \end{tabular}
    \label{tab:linelist}
\end{table}

\begin{table}[h]
    \centering
    \caption{Additional line list for abundance determination of $n$-capture elements in UVES spectra for DQ Hya. Hyperfine and isotopic splitting have been included for La II, Ba II and Eu II lines, as needed. We report the total oscillator strengths of the atomic transitions.}\label{tab:linelist_uves}
    \begin{tabular}{lccr}
    \hline
    Line & Species & E.P. (eV) & $\log gf$ \\
    \hline
    5087.416   & Y {\sc ii}   & 1.084 & -0.160  \\
    5200.406   & Y {\sc ii}   & 0.992 & -0.470 \\
    5289.815   & Y {\sc ii}   & 1.033 & -1.680 \\
    3991.127   & Zr {\sc ii}  & 0.758 & -0.310 \\
    4050.320    & Zr {\sc ii}  & 0.713 & -1.060   \\
    4211.877   & Zr {\sc ii}  & 0.527 & -1.040   \\
    4231.629   & Zr {\sc ii}  & 1.756 & -0.790  \\
    5112.270    & Zr {\sc ii}  & 1.665 & -0.850 \\
    4934.076   & Ba {\sc ii}  & 0.000 & -0.160  \\
    4921.786   & La {\sc ii}  & 0.244 & -0.450  \\
    6390.480   & La {\sc ii}   & 0.321 & -1.410  \\
    4882.463  & Ce {\sc ii}   & 1.528 & 0.190   \\
    5044.023  & Ce {\sc ii}   & 1.212 & -0.140  \\
    5075.355  & Ce {\sc ii}   & 1.026 & -0.350  \\
    5187.458  & Ce {\sc ii}   & 1.212 & 0.170   \\
    5274.229  & Ce {\sc ii}   & 1.044 & 0.130  \\
    5330.556  & Ce {\sc ii}   & 0.869 & -0.400   \\
    5393.392  & Ce {\sc ii}   & 1.103 & -0.060  \\
    5512.064  & Ce {\sc ii}   & 1.008 & -0.390    \\
    4020.860   & Nd {\sc ii}   & 0.321 & -0.190    \\
    4021.330   & Nd {\sc ii}   & 0.321 & -0.100   \\
    4825.480   & Nd {\sc ii}  & 0.182 & -0.420  \\
    5130.586   & Nd {\sc ii}  & 1.304 & 0.450   \\
    5255.502   & Nd {\sc ii}  & 0.205 & -0.670   \\
    5319.810   & Nd {\sc ii}  & 0.550 & -0.140   \\
    3819.670   & Eu {\sc ii}  & 0.000 & 0.490  \\
    4129.72   & Eu {\sc ii}   & 0.000 & 0.190  \\
    6645.064  & Eu {\sc ii}   & 1.380 & 0.421  \\
    \hline
    \end{tabular}
    \label{tab:linelist}
\end{table}


\subsection{Abundance sensitivities for DQ Hya}
We have calculated abundance sensitivities to stellar parameters for star DQ Hya. In Table \ref{tab:sensitivities} we report the resulting variation in [X/Fe] for all the species to changes in \teff, $\log g$ and \vmic.

\begin{table}[htbp]
    \centering
  \caption{Abundance sensitivities to change in stellar parameters for DQ Hya.}
    \begin{tabular}{l|cccr}
    \hline
      &    $\Delta$\teff & $\Delta\log g$ & $\Delta$\vmic  \\
      &         ($\pm 100K$) & ($\pm$ 0.2) & ($\pm 0.3$ km s$^{-1}$)  \\
      $\mathrm{\Delta[Fe/H]}$  & 0.07 & 0.01 & 0.04  \\
      $\mathrm{\Delta[C/Fe]}$  & 0.06 & 0.06 & 0.03 \\
      $\mathrm{\Delta[Mg/Fe]}$ & 0.08 & 0.02 & 0.06   \\
      $\mathrm{\Delta[Ca/Fe]}$ & 0.08 & 0.04 & 0.04  \\
      $\mathrm{\Delta[Ti/Fe]}$ & 0.05 & 0.03 & 0.05  \\
      $\mathrm{\Delta[Y/Fe]}$ & 0.08  & 0.11 & 0.04 \\
      $\mathrm{\Delta[Zr/Fe]}$ & 0.04 & 0.07 & 0.08  \\
      $\mathrm{\Delta[Ba/Fe]}$ & 0.09 & 0.03 & 0.24 \\
      $\mathrm{\Delta[La/Fe]}$ & 0.05 & 0.08 & 0.04 \\
      $\mathrm{\Delta[Ce/Fe]}$ & 0.09 & 0.07 & 0.02 \\
      $\mathrm{\Delta[Nd/Fe]}$ & 0.08 & 0.07 & 0.04  \\
      $\mathrm{\Delta[Eu/Fe]}$ & 0.07 & 0.07 & 0.04  \\
      \hline
    \end{tabular}
    \label{tab:sensitivities}
\end{table}

\subsection{Orbital properties}

In Table \ref{tab:orbitmemb}, we present a summary of the orbital properties: eccentricity, maximum height above the disc, and membership likelihood (equation \ref{eq:member}) to the thick disc ($p_\mathrm{thick}$), old thin disc ($p_\mathrm{thin, old}$; $7 < t_\mathrm{age}/\mathrm{Gyr} < 10$), intermediate-age thin disc ($p_\mathrm{thin, int}$; $3 < t_\mathrm{age}/\mathrm{Gyr} < 5$), and young thin disc ($p_\mathrm{thin, young}$; $t_\mathrm{age} < 3$ Gyr), as defined in \cite{robin2023}.
Values in parentheses indicate the membership likelihoods computed without considering the metallicity.
The reported values represent the mean over $10^5$ orbit realisations by sampling the uncertainties on the astrometric and chemical properties of the stars.  
Uncertainties are not shown, as they are below the $10\%$ level for the orbital parameters and $1\%$ level for the membership.
Different rows correspond to different Galactic potential models: \cite{mcmillan2017} (M17),
and \cite{allen1991} (AS91).
For DQ Hya, two variations are reported, corresponding to orbital integrations using the current distance estimated either from the parallax or from the absolute magnitude. For ASAS J153830–6906.4, only the parallax-based distance estimate is considered.

\begin{table*}[h!]
\centering
\caption{Orbital properties}
\begin{tabular}{lcccccr}
\hline
 Potential & ecc & \begin{tabular}[c]{@{}c@{}}$z_\mathrm{max}$\\ (kpc)\end{tabular} & $p_\mathrm{thick}$ & $p_\mathrm{thin, old}$ & $p_\mathrm{thin, int}$ & $p_\mathrm{thin,young}$ \\ \hline \hline
\multicolumn{7}{c}{DQ Hya (parallax distance)}  \\ \hline
M17  &   $0.12$  &   $1.90$   &      $0.99$ (0.66)     &  $0.01$   (0.17)       &      0  (0.17)      &       0    (0)        \\    
AS91     &  $0.11$  &   $1.90$   &      $0.99$ (0.64)     &  $0.01$   (0.18)       &      0  (0.18)      &       0    (0)     \\ \hline
\multicolumn{7}{c}{DQ Hya (magnitude distance)} \\ \hline
M17      &     $0.13$  &   $1.32$   &     $0.95$ (0.22)     &  $0.05$   (0.35)       &      0  (0.41)      &       0    (0.02)       \\
AS91     &  $0.12$  &   $1.31$   &     $0.94$ (0.21)     &  $0.06$   (0.35)       &      0  (0.42)      &       0    (0.02)       \\  \hline
\multicolumn{7}{c}{ASAS J153830-6906.4 }                        \\ \hline
M17      &    $0.24$  &   $0.61$   &     $1$ (0.11)     &  0  (0.40)       &      0  (0.47)      &       0    (0.02)         \\
AS91     &   $0.27$  &   $0.59$   &     $1$ (0.12)     &  0  (0.40)       &      0  (0.46)      &       0    (0.02)        \\ \hline
\end{tabular}
\label{tab:orbitmemb}
\end{table*}

\subsection{Spectral Energy Distribution for DQ Hya}

\begin{figure}[h!]
   \centering
   \includegraphics[width=0.55\textwidth]{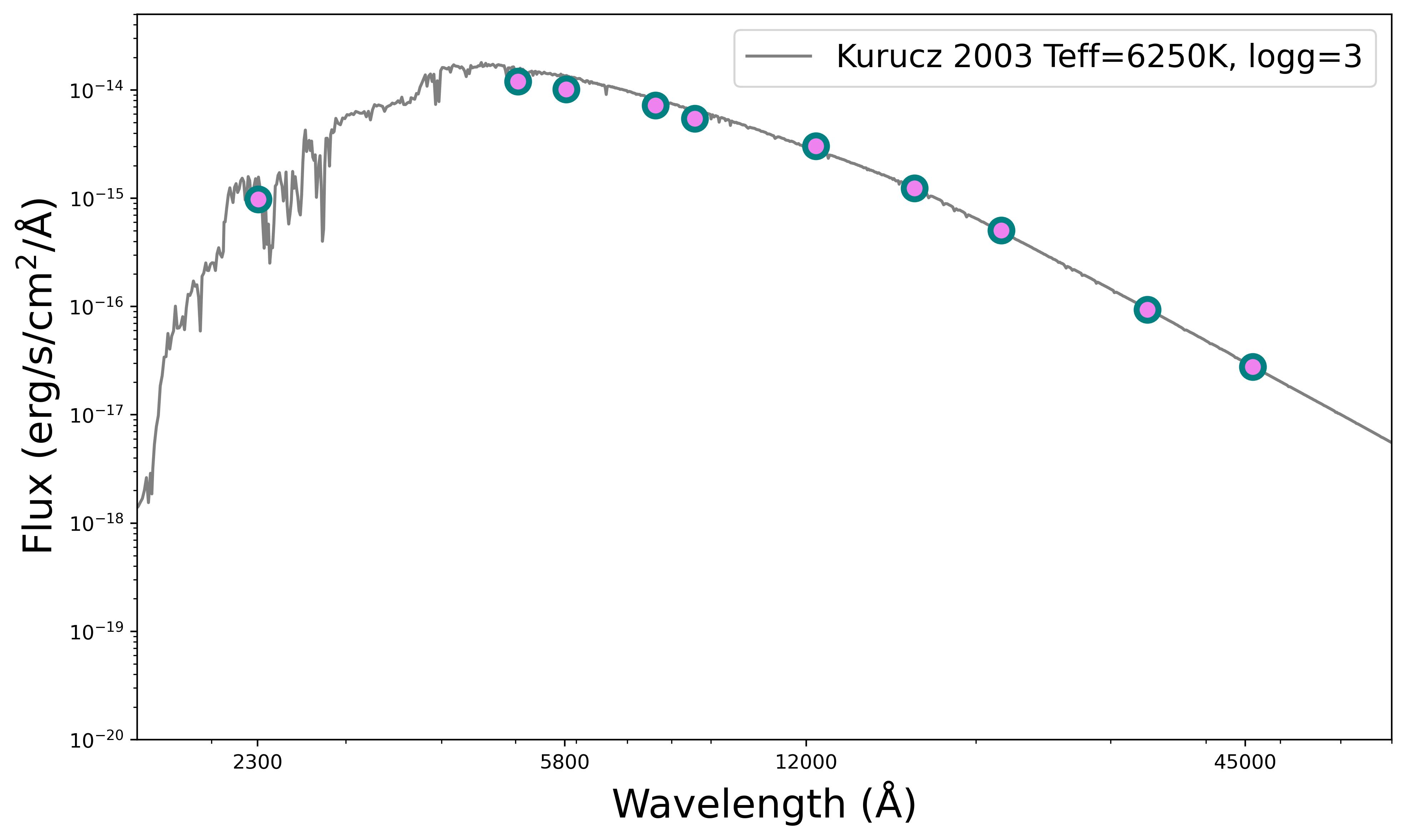}
      \caption{SED of DQ Hya, no excess in the GALEX NUV photometry is detected.}\label{fig:sed_dqhya}
   \end{figure}

In Figure \ref{fig:sed_dqhya}, we present the SED for star DQ Hya. Photometric data were obtained from GALEX \citep{martin2005}, Gaia DR3 \citep{gaiadr3}, 2MASS \citep{2mass}, and WISE \citep{wright2010}. No indication of NUV excess was observed when comparing the measured SED with the Kurucz \citep{castelli2003} atmospheric model corresponding to the star's parameters.

\end{appendix}

\end{document}